\documentclass[useAMS,usenatbib]{mnras}
\usepackage{graphics,epsfig,psfig}
\usepackage[normalem]{ulem}
\usepackage{xcolor}
\usepackage{float}
\usepackage{subfig}
\usepackage{graphicx}
\usepackage[]{inputenc,amssymb}
\usepackage{gensymb}
\usepackage{soul}
\usepackage{todonotes}

\def \be{\begin{equation}}
\def \ee{\end{equation}}
\def \bea{\begin{eqnarray}}
\def \eea{\end{eqnarray}}

\definecolor{webgreen}{rgb}{0,.5,0}
\definecolor{webbrown}{rgb}{.6,0,0}
\definecolor{cadmiumgreen}{rgb}{0.0, 0.42, 0.24}

\title[Imprint of patchy reionization on CMB]{Inevitable imprints of patchy reionization on the cosmic microwave background anisotropy}
\author[Paul, Mukherjee \& Choudhury]{
Sourabh Paul$^{1}$ \thanks{sourabh.paul@gmail.com},
Suvodip Mukherjee$^{2,3,4}$\thanks{ s.mukherjee@uva.nl}
\& 
Tirthankar Roy Choudhury$^5$ \thanks{tirth@ncra.tifr.res.in}\\
$^{1}$ Department of Physics and Astronomy, University of the Western Cape, Bellville, Cape Town, South Africa\\
$^{2}$ Gravitation Astroparticle Physics Amsterdam (GRAPPA),
Anton Pannoekoek Institute for Astronomy and Institute for High-Energy Physics,\\
University of Amsterdam, Science Park 904, 1090 GL Amsterdam, The Netherlands\\
$^{3}$ Institut d'Astrophysique de Paris,  98bis Boulevard Arago, 75014 Paris, France\\
$^{4}$ Sorbonne Universites, Institut Lagrange de Paris,  98 bis Boulevard Arago, 75014 Paris, France\\
$^{5}$ National Centre for Radio Astrophysics, Tata Institute of Fundamental Research, Pune 411007, India\\
}

\begin{document}
\label{firstpage}
\pagerange{\pageref{firstpage}--\pageref{lastpage}}
\maketitle

\label{firstpage}

\begin{abstract}
Reionization of the cosmic neutral hydrogen by the first stars in the Universe is an inhomogeneous process which produces spatial fluctuations in free electron density.  These fluctuations lead to observable signatures in cosmological probes like the cosmic microwave background (CMB). We explore the effect of the electron density fluctuations on CMB using photon-conserving semi-numerical simulations of reionization named \texttt{SCRIPT}. We show that the amplitude of the kinematic Sunyaev-Zeldovich (kSZ) and $B$-mode polarization signal depends on the patchiness in the spatial distribution of electrons along with the dependence on mid-point and extent of the reionization history. Motivated by this finding, we provide new scaling relations for the amplitude of kSZ and $B$-mode polarization signal which can capture the effects arising from the mean optical depth, width of reionization, and spatial fluctuations in the electron density arising from patchy reionization. We show that the amplitude of the kSZ and $B$-mode polarization signal exhibits different dependency on the width of reionization and the patchiness of reionization, and hence a joint study of these CMB probes will be able to break the degeneracy. By combining external datasets from 21~cm measurements, the degeneracy can be further lifted by directly exploring the sizes of the ionized regions. 
\end{abstract}

\begin{keywords} 
dark ages, reionization, first stars,  cosmic background radiation, cosmology: observations  
\end{keywords}

\section{Introduction}
The cosmic reionization of neutral hydrogen (HI) is a key epoch of pivotal importance that refers to the transition from the dark neutral phase to the present ionized Universe. Our current understanding suggests the beginning of this epoch to be at a redshift $z \sim 20 - 30$ when the first luminous sources in the Universe formed. The ultraviolet photons emitted by the first sources are capable of ionizing the neutral HI as the photoionization cross-section is very high at energies $\gtrsim 13.6$~eV. The neutral atoms in the proximity of the ionizing sources absorb these photons, which leads to bubbles of ionized hydrogen (HII). The contrast between neutral and ionized regions in the intergalactic medium (IGM) gives rise to large spatial fluctuation in the ionized fraction and thereby in the free electron density. This leads to the understanding that reionization has been a ``patchy'' phenomenon. The ionized bubbles expand, overlap and eventually merge completely, resulting in a uniformly reionized universe at $z\sim 6$ \citep{BarkanaLoeb:2001,BarkanaLoeb:2004,Furlanetto_2004,Wyithe_2003, 2018PhR...780....1D}.


It is now well accepted that cosmic reionization is a complicated phenomenon which depends on several astrophysical and cosmological effects. The studies of reionization at present attempts to answer several important open questions which can be listed as: (i) When did the first stars formed in the Universe? (ii) Is reionization driven by galaxies residing in rare massive haloes or abundant lighter haloes? (iii) How rapid is the process of reionization process? (iv) How inhomogeneous is the process of reionization? Understanding these issues requires accurate modelling of the physical processes during the EoR, which are unfortunately almost impossible to model from first principles. An alternate approach could be mostly data-driven where one attempts to reconstruct the EoR using multiple cosmological probes.  In this work, we attempt to understand how different characteristics of reionization (e.g., the timing, the extent, the sources responsible) affect different observational probes. 

One of the most important probes of the EoR has been the Cosmic Microwave Background (CMB) anisotropies. The so-called ``reionization bump'' in the Cosmic Microwave Background (CMB) anisotropy measurements of low $l$ $E$-mode polarization constrain the value of the reionization optical depth which results from the Thomson scattering of CMB photons through interaction with free electrons emerged from reionization. \citet{Planck:2018} put the latest constraints on the total optical depth to recombination at $\tau = 0.054 \pm 0.007$. At present, there are no direct observational constraints on the evolution of the ionization fraction during reionization, although theoretical models \citep{Mesinger:2007, Zahn:2007, Choudhury:2009} when compared with a variety of data, suggest that the reionization is not an instantaneous phenomenon, rather it is an inhomogeneous/patchy and complex process leading to spatial fluctuations in the ionized fraction during this epoch and future 21cm observations can potentially constrain the patchy reionization \citep[for recent reviews, see, e.g.,][]{2013ExA....36..235M,2015aska.confE..10M, 2016JApA...37...29C}.

Along with the large angular scale measurements of the CMB signal, observations from the ongoing and upcoming high-resolution ground-based CMB experiments (such as the Atacama Cosmology Telescope (ACTPol) \citep{2016ApJS..227...21T}, South Pole Telescope (SPT) \citep{Benson:2014qhw}, Simons Observatory \citep{Ade:2018sbj}, and CMB-S4 \citep{Abazajian:2019eic}), can also explore the small angular scale anisotropies in CMB temperature and polarization field. Proposed CMB mission concepts (such as Probe of Inflation and Cosmic Origins (PICO) \citep{Hanany:2019lle}, CMB-Bharat \footnote{\url{http://cmb-bharat.in/}}, CMB-HD \citep{Sehgal:2020yja}, and the proposal submitted to Voyage-2050 \citep{Delabrouille:2019thj}) are also capable of probing secondary anisotropies in the CMB with high angular resolution. The small angular scale CMB anisotropies are a rich source of information about secondary CMB anisotropies which are generated after the surface of the last scattering, which is around $z \sim 1080$. 
The CMB photons that we observe today have interacted with matter along their path and result in the secondary anisotropies. The interactions comprise of gravitational lensing \citep{Seljak:1997ep, Zaldarriaga:1998ar, Hu_2001, Hu_2002}, the thermal Sunyaev-Zeldovich effect \citep[tSZ; the inverse Compton scattering of CMB photons with free electrons,][]{Sunyaev_1970}; and the kinematic Sunyaev-Zeldovich effect \citep[kSZ; Doppler effect due to bulk motion of free electrons in, e.g., galaxy clusters and HII bubbles, which can Thomson scatter the CMB photons,][]{Sunyaev_1980,Nozawa:1998}. 

During reionization, the contribution of kSZ to the secondary anisotropies exceeds greatly compared to tSZ \citep{Aghanim_2008}. The influence of patchy reionization on small scale CMB anisotropies have been widely studied using analytical models and numerical simulations \citep{Knox:1998fp,1998ApJ...508..435G, Santos_2003, Zahn_2005, Mcquinn_2005, 2005MNRAS.360.1063S, Mesinger_2012, Battaglia_2013, Park_2013, Calabrese:2014gwa,Alvarez_2016, Gorce:2020pcy}. High-resolution CMB experiments have also provided observational evidence of the kSZ signal originated during reionization \citep{Zahn_2012,2012ApJ...755...70R,Dunkley:2013vu,Sievers:2013ica, Crawford:2013uka, 2020arXiv200206197R}.
The patchy reionization also gives rise to secondary $B$-mode polarization and leaves an imprint on the measurements of the primordial $B$-mode polarization \citep{Hu:2000,Santos_2003,Mortonson:2007, Dvorkin:2008tf, Dvorkin:2009, 2011arXiv1106.4313S}. In a previous paper \citep{Mukherjee_2019}, we have studied this effect in detail using simulations and estimated the patchy reionization contamination in the primordial $B$-mode polarization signal.

In this work, we focus on observational probes based on CMB to understand how they can be used to constrain the patchiness in EoR and hence infer about the sources which drove the process. Our main focus is on the secondary anisotropies in the temperature and polarization field of the CMB  such as kSZ temperature fluctuations, and $B$-mode polarization signal (even in the absence of primordial gravitational waves). We also complement these probes of the CMB with the temperature brightness fluctuations in the redshifted 21~cm signal of HI, which can be useful in relating the angular scales of the observed signal to physical quantities like the size of ionized regions.
 
The plan of the paper is as follows: we further elaborate on the motivation of this work in section~\ref{Motivation}. In section~\ref{Theory}, we provide a theoretical overview of the kSZ effect, the patchy reionization contribution to the secondary CMB $B$-mode polarization and the Cosmological 21cm signal from the reionization. The details of our semi-numerical simulation prescription for generating the patchy reionization signal is given in section~\ref{Simulation}. In section~\ref{Results}, the main outcomes of our work are discussed in detail and finally, section~\ref{Conclusion} contains the primary findings of this paper and scopes of future work. Throughout this paper, we have used the flat $\Lambda$CDM cosmological parameters $[\Omega_m, \Omega_b, h, n_s, \sigma_8] = [0.308, 0.0482, 0.678, 0.961, 0.829]$ from \citet{Planck:2014} which is consistent with \citet{Planck:2018}.

\section{Motivation: Connecting the physics of EoR with cosmological observables}
\label{Motivation}

The presence of spatial fluctuations in the electron density during the EoR leads to detectable signal in the CMB anisotropies, namely, the kSZ power spectrum and the $B$-mode polarization power spectrum. Conventionally, it is assumed that the kSZ signal is mostly determined by the mean reionization redshift and the extent of reionization \citep{Mcquinn_2005, Zahn_2012, Battaglia_2013}. However, in addition to the contribution arising from the width of reionization,  the contributions to the kSZ signal depend on the total amount of integrated fluctuations in the momentum field during the EoR. As a result, it is expected that the kSZ amplitude will depend not only on the width of the reionization and the mean redshift but also on the extent of spatial fluctuations present in the electron density and thereby the patchiness during the EoR \citep{Park_2013,Gorce:2020pcy}.

For a measured value of the optical depth $\tau$, reionization starting late and taking place rapidly, can be driven by massive halos and hence can generate more fluctuations in electron density (see Sec. \ref{Simulation}). This can lead to high kSZ signal due to the contribution from the momentum field even if the width of the reionization $\Delta z$ is shallow. In the opposite end, a reionization happening slowly with a large value of $\Delta z$, can be driven by small halos and can have less fluctuation in the electron density in the early EoR, and a significant contribution can arise from the wider value of $\Delta z$. Along with these two effects, change in the value of the optical depth $\tau$ (or the mean redshift $\bar z$) can alter the strength of the signal. As a result, the total signal in kSZ during the EoR should be an interplay between the value of $\tau$, $\Delta z$, and spatial fluctuation in electron density. 

Similar to the kSZ signal, the secondary anisotropy due to Thomson scattering of CMB quadrupole by the free electrons (distributed inhomogeneously) produces $B$-mode polarization signal. The strength of the signal depends on the integrated electron density fluctuation during the EoR. Reionization driven by massive halos can lead to large spatial fluctuations in electron density \citep{Mukherjee_2019}. Also, for a longer (or shallower) duration of reionization, the net integrated contribution to $B$-mode polarization should be more (or less). As a result, similar to the kSZ signal, secondary $B$-mode polarization should also depend on the value of $\tau$, $\Delta z$, and spatial fluctuation in electron density.

In this paper, we have obtained a new scaling relation for kSZ signal and $B$-mode polarization signal as given in equations (\ref{newscaling}) and (\ref{newscalingbmode}), respectively which also captures the contribution from spatial fluctuation in electron density. In contrast to other works, our scaling relation is written in terms of the mean optical depth $\tau$ rather than in terms of the mean value $\bar z$. We have considered this route because the former one is a directly observable quantity (from small scale temperature fluctuations in CMB and large scale polarization fluctuations in $E$-mode polarization), whereas the latter one is model-dependent and depends on the assumed reionization history. These new scaling relations make it possible to connect the CMB observables (kSZ, $B$-mode polarization) with the characteristics of reionization.

\section{Overview of the theoretical formalism}
\label{Theory}
\subsection{kSZ during the EoR}
The bulk motion of the ionized bubbles with respect to the CMB gives rise to the kSZ effect and leads to secondary anisotropies in CMB temperature field\footnote{The choice of the direction is taken such as the bubble which is moving towards us leads to an excess temperature fluctuation.}
\begin{equation}
\frac{\Delta T (\hat{n})}{T_0} = -\int_0^{\tau} d\tau ~e^{-\tau(\chi)} \frac{\hat{n}\cdot {\bf v}}{c},
\label{ksz}
\end{equation} 
where $T_0 = 2.725$K is the CMB temperature at $z=0$, $\hat{n}$ denotes the line of sight unit vector and {\bf v} refers to the peculiar velocity field. The quantity $\tau(\chi)$ is the integrated Thomson scattering optical depth through the IGM from present epoch to the redshift of interest:
\begin{equation}
\tau(\chi) = \sigma_T \bar{n}_{H} \int_0^{\chi} d\chi~ (1+z)^2 x_e(1+\delta),
\label{tau}
\end{equation}
with $\tau$ being the integrated optical depth to the last scattering surface. In the above equation, $\sigma_T$ is the Thomson-scattering cross section, $\chi$ indicates the comoving distance to the epoch of interest and $\bar{n}_{H}$ is the \emph{comoving} mean number density of hydrogen. The quantities $\delta$ and ${\bf v}$, respectively,  are the density contrast and peculiar velocity of baryons. We denote the free electron fraction as $x_e = \chi_{\rm{He}} x_{\rm{HII}}$ with $\chi_{\rm{He}} = 1.08$ referring to the excess electron correction factor due to singly ionized Helium and $x_{\rm{HII}}$ being the ionized fraction of hydrogen. Therefore, the global mass-averaged ionization fraction is given by by $Q_{\rm{HII}}(z) \equiv \langle x_{\rm{HII}} (1+\delta)\rangle$.

As $\tau$ is directly proportional to the free electron number density, we can define a dimensionless ionized momentum field as
\begin{equation}
{\bf q} = x_e(1+\delta) \frac{{\bf v}}{c}.
\label{q}
\end{equation}
In terms of {\bf q}, equation~(\ref{ksz}) can be written as: 
\begin{equation}
\frac{\Delta T (\hat{n})}{T_0} = - \sigma_T \bar{n}_{H}\int \frac{d\chi}{a^2}e^{-\tau(\chi)} {\bf q}\cdot\hat{n}.
\label{ksz1}
\end{equation}

The kSZ angular power spectrum can be estimated from the momentum field and calculated using Limber's approximation \citep{Limber_1953} as \citep{Ma_Fry_2002, Park_2013, Alvarez_2016} :
 \begin{equation}
C_l^{\rm{kSZ}} = \left(\sigma_T \bar{n}_{H}T_0\right)^2 \int \frac{d\chi}{\chi^2 a^4}e^{-2\tau(\chi)}\frac{P_{q_\perp}(k = l/\chi, \chi)}{2}.
\label{Cl_eq}
\end{equation}
Here ${\bf q_\perp}$ is the transverse component of the ${\bf q}$ field, i.e., its direction is perpendicular to the ${\bf k}$ vector (${\bf k} \cdot {\bf q_\perp} = 0$) and hence can be calculated as ${\bf q_\perp(k)} = {\bf q(k)} - [{\bf q(k)} \cdot \hat{k}]\hat{k}$ \citep{Park_2013}. $P_{q_\perp}(k, \chi)$ is the power spectrum of transverse component and is defined as $\langle {\bf q_\perp}({\bf k}, \chi) \cdot {\bf q_\perp}^*({\bf k'}, \chi)\rangle = (2\pi)^3 P_{q_\perp}(k, \chi) \delta_D({\bf k}-{\bf k'})$. 

In reality, the kSZ is an integrated effect that gets contribution from both during and post reionization epochs. During post-reionization, the signal is sourced by the Ostriker-Vishniac (OV) effect \citep{1986ApJ...306L..51O, Ma_Fry_2002}, which requires modelling of the non-linear density and velocity fields \citep{2012ApJ...756...15S}. We do not attempt to model the OV effect in this paper and therefore set the lower integration limits for equations~(\ref{ksz1}) and (\ref{Cl_eq}) to the redshift $z_{\rm end}$ where reionization is complete (which depends on the reionization model under consideration). This implies that the kSZ signal amplitudes quoted in the paper take into account only the contribution from patchy reionization.

\subsection{Secondary CMB polarization during the EoR}

In addition to \sout{small-$l$} temperature anisotropies, the spatial fluctuations in the electron density during the EoR leads to secondary fluctuations in the polarization field of CMB. The signal arises due  to two effects which are called scattering and screening \citep{Hu:2000,Santos_2003,Mortonson:2007, Dvorkin:2008tf, Dvorkin:2009, 2011arXiv1106.4313S}. Scattering arises due to the Thomson scattering of the CMB photons (with quadrupole anisotropy) with the inhomogeneous spatial distribution of electrons during the EoR, which can be written as\footnote{This expression is valid at all angular scales {under the approximation} of constant source (or slowly varying source \cite{Hu:2000}).} 
\begin{eqnarray}\label{bb-power-exact}
C_l^{BB, sca}&=& \frac{24\pi \bar{n}^2_{H}\sigma^2_T}{100} \int d\chi \frac{1}{a^2}\int d\chi' \frac{1}{a'^2}e^{-\tau(\chi)- \tau(\chi')} 
\nonumber \\
&\times& \int dk \frac{k^2}{2\pi^2} P_{ee} (k, \chi, \chi') j_l(k\chi)j_{l}(k\chi') \frac{Q_{\rm RMS}^2}{ {2}},
\nonumber \\
\end{eqnarray}
where the power spectrum of the electron density fluctuations $\Delta_e \equiv x_e (1 + \delta)$ is written as $\langle \Delta_e ({\bf k},\chi')  \Delta_e^* ({\bf k'},\chi')\rangle  \equiv P_{ee} (k, \chi, \chi') \delta_D({\bf k} - {\bf k}')$. The parameter $Q_{\rm RMS}^2$ denotes the quadrupole temperature variance, which is considered to be $22 \, \mu$K over the redshift range of reionization and $j_l(k\chi)$ are the spherical Bessel functions. This signal is dominant at large angular scales and can be a  source of contamination to the primordial $B$-mode polarization signal \citep{Mortonson:2007, Mukherjee_2019, Roy:2020cqn}. For angular scales $l\gtrsim30$ \citep{Mukherjee_2019}, the $B$-mode power spectrum can be estimated using the Limber's approximation \citep{Limber_1953} as:
\begin{eqnarray}\label{bb-power-limber}
C_l^{BB, sca}&=& \frac{6 \bar{n}^2_{H}\sigma^2_t}{100} \int  \frac{d\chi e^{-2\tau(\chi)}}{a^4\chi^2}
P_{ee} \left(k=\frac{l+1/2}{\chi}, \chi\right)\nonumber \\
&\times& \frac{Q_{RMS}^2}{ {2}}.
\end{eqnarray}
In our previous study \citep{Mukherjee_2019}, we used semi-numerical simulations of the EoR to show that there can be up to $30\%$ bias in the value of tensor to scalar ratio $r=0.001$ measured by the upcoming CMB probes of the $B$-mode polarization and can also lead to an increase in the error-bar on $r$. 

The secondary effect such as screening arises from the rotation of the primordial CMB polarization field due to the inhomogneous optical depth along the line of sight. This effect is dominant only at small angular scales over the scattering effect, and can be written under the flat-sky approximation as \citep{Dvorkin:2009, 2013PhRvD..87d7303G}
\begin{eqnarray}
C_l^{BB, scr}&=& \int \frac{d^2{l'}}{(2\pi)^2} C^{EE}_{l'} C^{\tau\tau}_{|l-l'|}\sin{2\phi_{l'}}
\label{bb-power-exact}
\end{eqnarray}
where, $C^{EE}_l$ and $C_l^{\tau\tau}$ are the angular power spectrum of the primordial $E$-mode polarization signal of CMB and optical depth $\tau$ due to inhomogeneous cosmic reionization.  

\subsection{Cosmological 21~cm signal during the EoR}

Although the main focus of this work is to understand how the physics of patchy reionization affects the CMB anisotropy signals, we also study the 21~cm signal predicted by the models. Since the CMB signal is an integrated effect along the line of sight, often the details of reionization get integrated out and thus makes it difficult to understand. Since the 21~cm signal follows the EoR at individual redshifts, it helps to compare the CMB signals with the 21~cm predictions. It is worth mentioning here the major experimental efforts that are being undertaken to study the EoR using redshifted 21~cm emission. In particular, radio interferometers such as the Low Frequency Array \citep[LOFAR,][]{Haarlem_2013}, Giant Meterwave Radio Telescope \citep[GMRT,][]{Paciga_2013}, Donald C. Backer Precision Array for Probing the Epoch of Reionization \citep[PAPER,][]{Parsons_2014}, and the Murchison Widefield Array \citep[MWA,][]{tingay_2013} are currently operational in the $80<\nu<300$ MHz range to detect the redshifted 21~cm line. In the future, the Hydrogen Epoch of Reionization Array \citep[HERA,][]{DeBoer_2017} and SKA1-low\footnote{\url{https://www.skatelescope.org}} will be operational with increased sensitivity for both direct and statistical detection. 

The differential brightness temperature of the predicted 21~cm signal from EoR with the CMB as background is given by \citep{Field:1958, Field:1959, Furlanetto_2004}:
\begin{eqnarray}
\delta T_b({\bf x},z) &=& 27~\rm{mK}~x_{\rm{HI}}({\bf x},z)[1+\delta({\bf x},z)]
\nonumber \\
&\times& \left(\frac{\Omega_B h^2}{0.023}\right)  \left(\frac{0.15}{\Omega_m h^2} \frac{1+z}{10}\right)^{1/2},
\label{21cm}
\end{eqnarray}
where $x_{\rm{HI}} \equiv 1 - x_{\rm HII}$ is the neutral hydrogen fraction and $T_\gamma (z) = 2.725(1+z)$ is the CMB temperature. The above relation assumes that the spin temperature of neutral hydrogen is much larger than $T_{\gamma}$ which is valid because of the X-ray heating from the early sources.


\section{Semi-numerical simulations of cosmic reionization using \texttt{SCRIPT}}
\label{Simulation}
In this section, we describe the model of reionization we have used to generate the kSZ and 21 cm signals. We have used the Cosmological $N$-body simulation code GADGET-2 \citep{GADGET2:2005} to generate the large-scale dark matter density field on a 512$h^{-1}$Mpc size box containing $256^3$ collisionless dark matter particles. Assuming that the baryon density field at high-$z$ trace the dark matter distribution, snapshots were generated for $5\leq z \leq 15$ at equal redshift interval $\delta z = 0.1$. We assume that each halo above some minimum mass $M_{\rm{min}}$ is capable of ionizing a mass of intergalactic hydrogen atoms that is proportional to the halo mass.

Due to the low resolution of our simulation setup and the choice of minimum halo mass, the traditional halo finder algorithms are not able to resolve collapsed haloes responsible for producing the ionizing photons. Although, low mass haloes are highly abundant in the initial stage of reionization and act as major drivers for early reionization of the Universe. Therefore, in order for the resulting ionization field to reflect the underlying density field correctly, we employ an approach based on conditional mass fraction due to ellipsoidal collapse to incorporate the contribution of small mas haloes which are not resolved otherwise \citep{2002MNRAS.329...61S, Seehars:2015ada, Choudhury:2018}. The quantity of interest here is the collapsed fraction $f_{\rm{coll}}(M_{\rm{min}})$ which denotes the mass fraction in a given volume that resides in haloes of mass $M_{\rm{min}}$ or higher. 

As the end product of this prescription, we have the collapsed fraction $(f_{\rm{coll}})$ and value of density fluctuation for every grid cell in the simulation box. To construct the ionization field from here, we use the photon-conserving semi-numerical scheme introduced in \citet{Choudhury:2018}, which is named \texttt{SCRIPT} (\textbf{S}emi-numerical \textbf{C}ode for \textbf{R}e\textbf{I}onization with \textbf{P}ho\textbf{T}on-conservation). In addition to solving the non-conservation of photons in excursion set bases models \citep{Mesinger:2007, Zahn:2007, Geil:2007rj, Santos:2007dn, Choudhury:2009, 2011MNRAS.411..955M}, it also ensures the numerical convergence of large-scale power spectrum of the ionization field with respect to the resolution at which the method is employed. 

The choice of box size plays a crucial role in estimating the kSZ power spectrum. Small box sizes (100 Mpc/h or less) fail to capture large-scale velocity modes and therefore inaccurately depicts the kSZ fluctuation \citep{2012ApJ...756...15S}. For the post-reionization kSZ signal, it is relatively straightforward to account for the missing longer wavelengths from the linear perturbation theory. For patchy reionization, \citet{Park_2013} corrects for the missing power in a finite simulation box size with the assumption that the density and ionization fields are relatively flat on large scales. They apply the velocity correction on a smaller box size of 114$h^{-1}$Mpc and find good agreement with a larger box size of 425$h^{-1}$Mpc. In this paper, we choose the box size of 512$h^{-1}$Mpc for faster numerical calculations without sacrificing too much of the kSZ power from large-scale velocity modes. Moreover, for a given $l$, the modes that contribute to the kSZ power fall within a small $k$ range which can be approximated as $k \sim\, l/\chi$. For $l=3000$, this corresponds to the $k$ range of $0.42\, h$/Mpc$-0.54\, h$/Mpc for the redshift range of our interest z $\in\, 5-15$ i.e. reionization. Therefore, our choice of the box size captures almost all the required velocity modes and provides robust estimates of the kSZ power at this scale.

\subsection{Generating kSZ and 21 cm light cones}

The simulation along with the steps discussed above deliver us the so-called ``coeval cubes'' which are three-dimensional volumes of density fluctuation, radial velocity and ionization fraction fields at each cosmological redshift. Obviously, the cartesian dimension of the three-dimensional volume does not represent the redshift axis as all elements in a single cube represent simultaneous evolution and reflect the IGM at a fixed redshift. For example, if we place the center of a 512$h^{-1}$Mpc box at the comoving distance corresponding to redshift $z = 8$, then the sides nearest and furthest from the observer represent the redshifts $z \approx 6.9$ and $z \approx 9.4$ (for the flat $\Lambda$CDM cosmology used in this paper). This means that such a scenario indicates no redshift evolution between $z=6.9$ and $9.4$. Therefore, light cone volumes  must be generated from the ``coeval cubes'' to characterize the evolution of IGM along the line of sight and illustrate redshift evolution \citep{BarkanaLoeb:2006, Datta:2012, LaPlante:2014, Ghara:2015, 2018MNRAS.474.1390M}. In this work, we follow the prescription outlined in \citet{Ghara:2015} to generate the light cones.


With the light cone cubes constructed for density fluctuation, radial velocity and ionization fraction; equations~(\ref{tau})--(\ref{ksz1}) and (\ref{21cm}) are used to compute the kSZ and 21~cm brightness temperature maps. Note that these light cones are generated only for visualizing the evolution of the EoR and the main effects which determine the signal; however, these are \emph{not} required for calculating any of the CMB signals.

\section{Results from the simulations}
\label{Results}

\begin{figure*}
\centering
\includegraphics[trim={0cm 0cm 0cm 0cm},clip,width=1.0\textwidth]{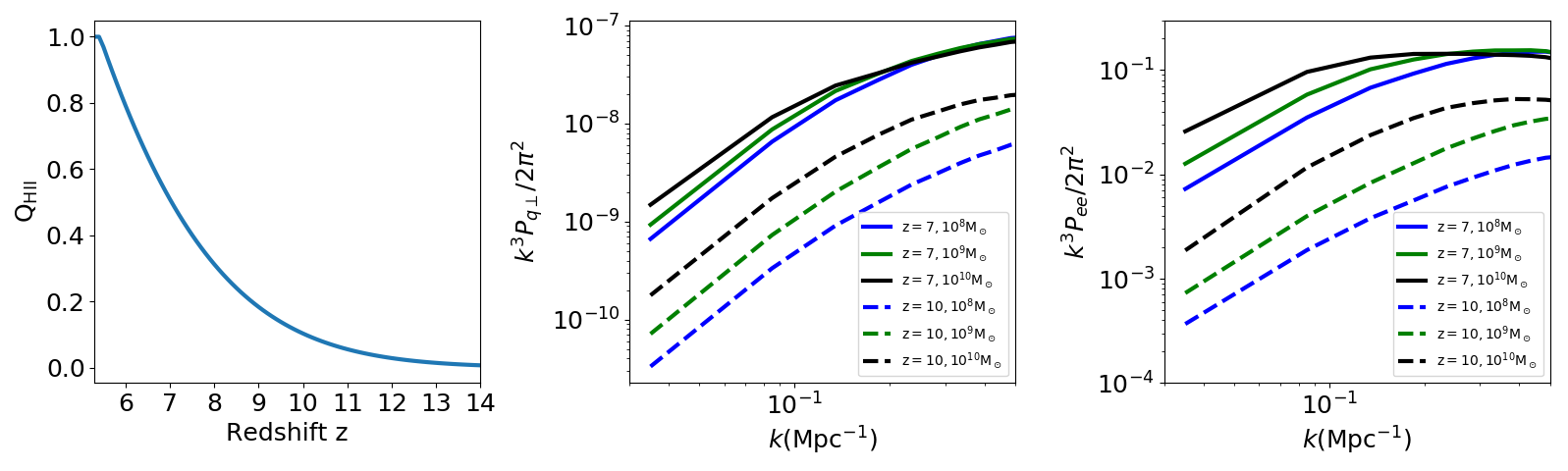}
\caption{The left panel shows the evolution of the ionized mass fraction $Q_{\rm HII}(z)$ for our fiducial reionization history. The middle panel shows the  dimensionless power spectra of the transverse component of the momentum field $({\bf q_\perp})$ at redshifts $z=7$ and $10$ for ${\rm{M_{\rm min} = 10^8, 10^9}}$ and $\rm{10^{10} M_\odot}$ using \texttt{SCRIPT} keeping the reionization history same as the fiducial model. The dimensionless power spectra of the electron distribution power spectrum $P_{ee}$ for the same cases are shown in the right panel.}
\label{fiducial}
\end{figure*}

\begin{figure*}
\centering
{\includegraphics[trim={3cm 0.5cm 2.7cm 0.5cm},clip,width=0.8\textwidth]{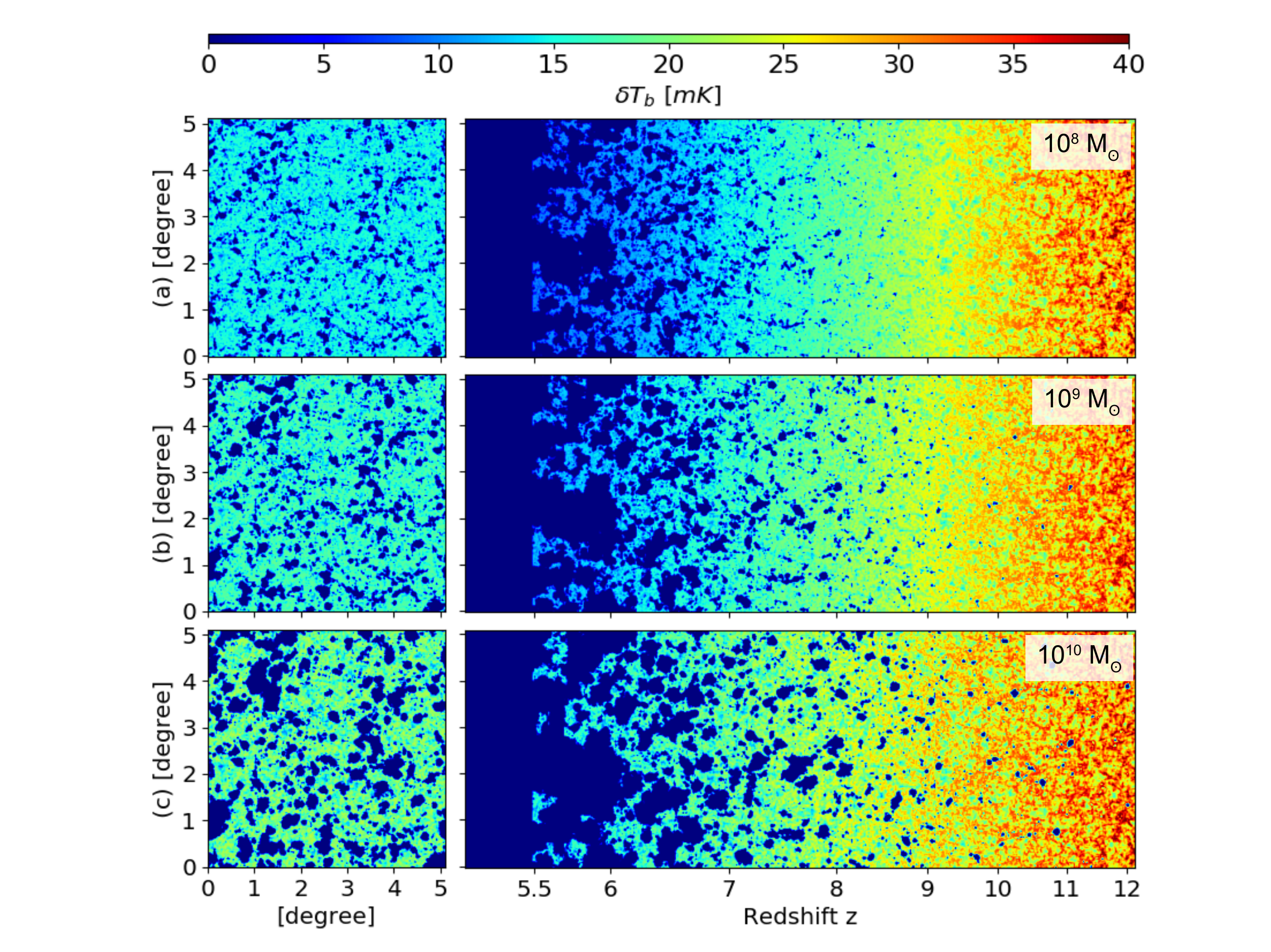}}
\caption{Visualization of the 21cm differential brightness temperature $\delta T_b$ for a single realization. The top, middle and bottom rows correspond to the three minimum halo mass cases considered in this paper: $M_{\rm{min}}=10^8, 10^9$ and $10^{10} \rm{M}_\odot$ respectively. Left panel: a single slice from the coeval cube at approximately half-reionization redshift $z=7$ with $Q_{\rm{HII}}=0.51$. Right panel: the evolution of the 21cm signal across redshift in the form of a light cone cube. The signal follows the underlying matter fluctuation at the initial stage of reionization and gradually disappears at lower redshifts as the IGM gets ionized. For higher minimum halo mass cases, both left and right panel portray larger bubble sizes with distinctive boundaries which in turn increases the patchiness in the IGM.}
\label{21cm_cones}
\end{figure*}

\begin{figure*}
\centering
{\includegraphics[trim={3cm 0.5cm 2.7cm 0.5cm},clip,width=0.8\textwidth]{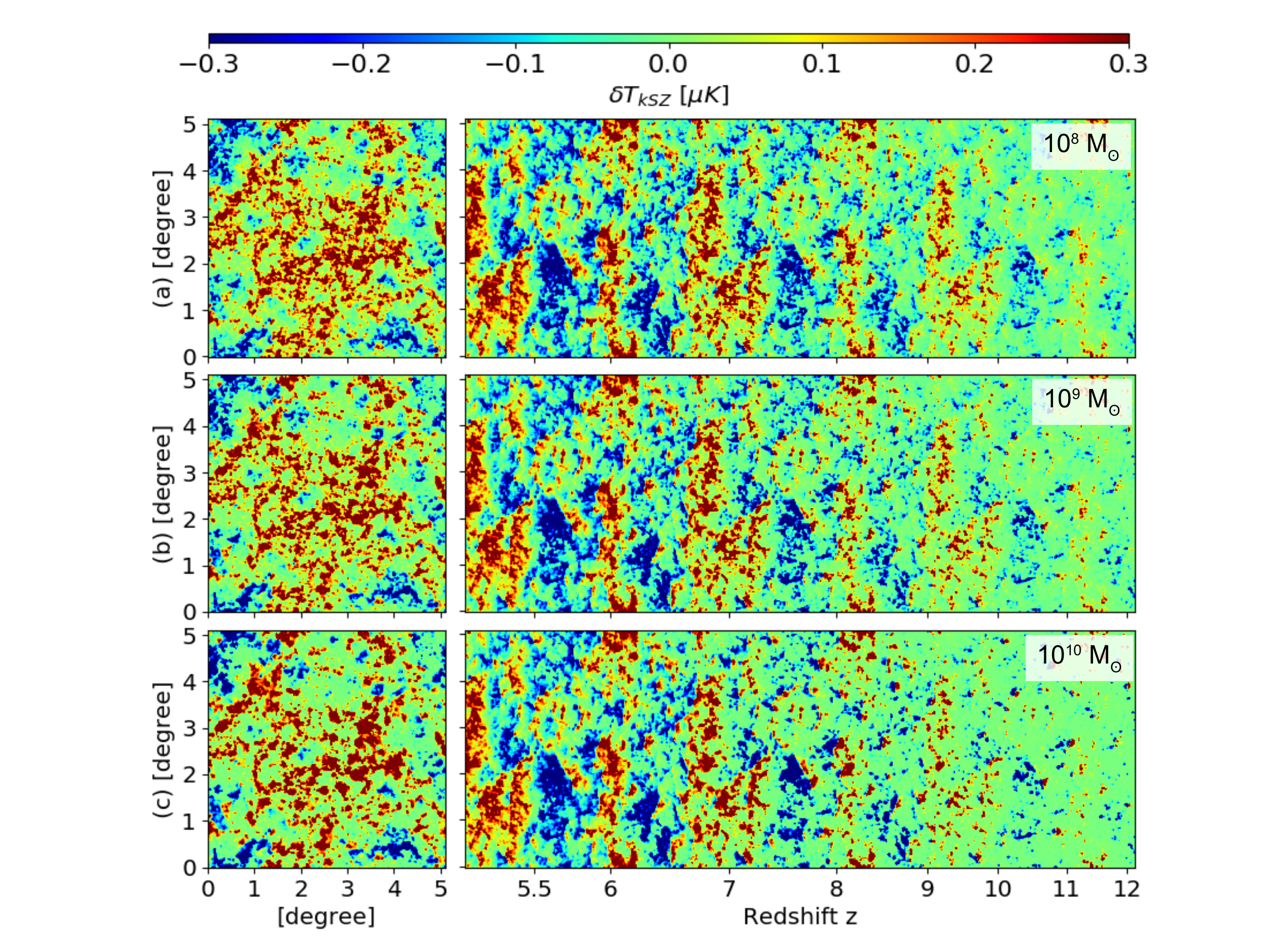}}
\caption{Visualization of the kSZ signal for the same realization as presented in Figure~\ref{21cm_cones} for ${\rm{M_{\rm min} = 10^8, 10^9}}$ and $\rm{10^{10} M_\odot}$ (from top to bottom) for the fiducial reionization case. The left panel shows the kSZ contribution from the slice at $z = 7$ with 2Mpc/h width. The right panel describes the evolution of the KSZ signal across redshift.}
\label{ksz_cones_M}
\end{figure*}

\subsection{The fiducial reionization history}

To construct a fiducial model of reionization, we assume that all haloes with mass $\geq M_{\rm min} = 10^8 {\rm{M}}_\odot$ are able to ionize the IGM. This choice of halo mass is driven by the corresponding virial temperature of $10^4$K for efficient cooling mediated by atomic transition, which is required for fragmentation of primordial gas into stars when they fall into dark matter haloes \citep{BarkanaLoeb:2001}. We assume the ionizing efficiency $\zeta$ to be $z$-independent, and we fix the value of $\zeta$ in a way that the resulting reionization history (left-hand panel of Figure~\ref{fiducial}) results in $\tau = 0.054$, the mean value from the latest Planck results \citep{Planck:2018}.

We also consider two more reionization scenarios by increasing $M_{\rm{min}}$ to $10^{9}$ and $10^{10}{\rm{M}}_\odot$. The value $M_{\rm{min}} = 10^{9}{\rm{M}}_\odot$ corresponds to the cases where radiative feedback suppresses star formation in low-mass haloes \citep{2008MNRAS.385L..58C}, whereas the extreme case of $M_{\rm{min}} = 10^{10}{\rm{M}}_\odot$ is motivated by AGNs driven reionization mechanism \citep{Kulkarni:2017qwu}. From hereon, the cases of $M_{\rm{min}} = 10^{8}, 10^{9}$ and $10^{10}{\rm{M}}_\odot$ are denoted as M8, M9 and M10 respectively. For M9 and M10, $\zeta$ is varied across redshift so that the resulting $Q_{\rm{HII}}(z)$ is consistent with the reionization history of M8 (left hand panel of Figure~\ref{fiducial}). 

In the middle panel of Figure~\ref{fiducial}, we show the dimensionless power spectrum $P_{q_{\perp}}(k)$ of the transverse component of the momentum field $({\bf q_\perp})$ at two redshifts $z=7$ and $z = 10$ for ${\rm{M_{\rm min} = 10^8, 10^9}}$ and $\rm{10^{10} M_\odot}$ for the fiducial reionization history. We find that the power spectrum amplitude is higher at $z = 7$ than at $z = 10$. This is because the bubble distribution is significantly more patchy at $z = 7$ where $Q_{\rm HII} \sim 0.5$. We also find that the power spectrum amplitude increases with increasing $M_{\rm min}$. This is also along the expected lines as the reionization is driven by relatively rarer sources when $M_{\rm min}$ is larger, leading to relatively fewer but large-sized bubbles, thus leading to more fluctuations. We also show the dimensionless power spectrum $P_{ee}(k)$ of the electron fraction for the same cases in the right-hand panel of Figure~\ref{fiducial}. The observations, in this case, are almost identical to the case of $P_{q_{\perp}}(k)$.

To understand this further, we show the light cones for the 21 cm brightness temperature field in the right panels Figure~\ref{21cm_cones} for the three cases M8, M9 and M10. In the left panels of the same figure, we show a single slice of the coeval boxes for M8, M9 and M10 at $z=7$ ($Q_{\rm{HII}} = 0.51$). Note that all the rows have the same reionization history, the only difference being the sources of reionization. Clearly, larger values of $M_{\rm min}$ leads to larger (but fewer) ionized bubbles and hence more contrast in the ionization field. Hence, we expect that any signal that depends on the fluctuations in the ionization field to be larger for higher $M_{\rm min}$.

\begin{figure}
\centering
{\includegraphics[trim={0cm 0.2cm 0cm 0cm},clip,width=0.45\textwidth]{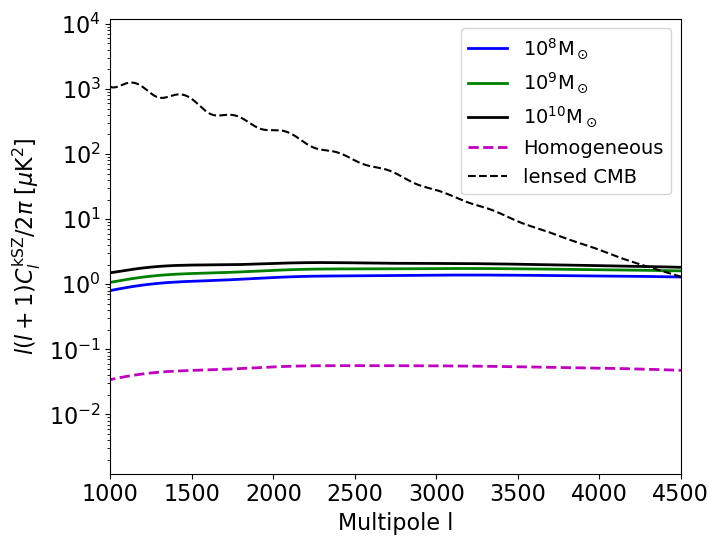}}
\caption{kSZ angular power spectrum estimated using equation~(\ref{Cl_eq}) which applies the Limber's approximation to calculate $C_l^{\rm{kSZ}}$ from 3D transverse momentum field power spectra $P_{q_{\perp}}$ integrated over redshift. The lensed angular power spectrum of CMB temperature fluctuations are obtained using CAMB \citep{Lewis:1999bs, 2012JCAP...04..027H} to compare the amplitude with kSZ. The plot indicates high-$l$ modes are more powerful to explore the kSZ signal.}
\label{kSZ_Cl}
\end{figure}

For comparison, we also show the light cones for the kSZ temperature field in the right panels Figure~\ref{ksz_cones_M}, while the left panels show a single slice of the coeval boxes at $z = 7$. Note that the observable kSZ signal is an integral over the light cone. The first point to note is that, unlike the 21~cm signal, the kSZ signal continues to exist even when reionization is over (i.e., at $z < 5.4$). This is simply due to the density fluctuations and the velocity field and is ignored in our work. In the EoR, we find that the ionized bubbles (blue regions) are larger for higher $M_{\rm min}$, thus leading to higher amplitude in the kSZ signal.

\subsection{Impact of reionization sources on the kSZ angular power spectrum}

\begin{figure*}
\centering
{\includegraphics[trim={0cm 0cm 0cm 0cm},clip,width=1.0\textwidth]{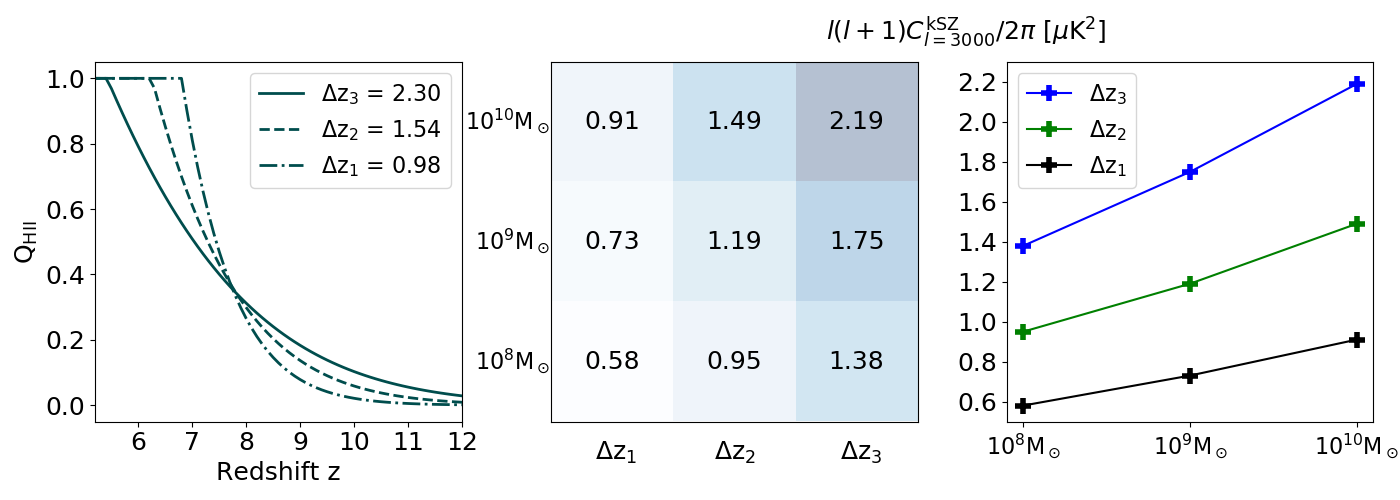}}
\caption{Dependence of kSZ angular power spectra on various reionization parameters. Left panel: three reionization cases which are distinguished by $\Delta z$; all cases result in $\tau \approx 0.054$. Middle panel: 2d distribution of $l(l+1)C_{l=3000}^{\rm{kSZ}}/2\pi$ on various minimum halo mass and $\Delta z$ cases. Right panel: Effect of patchiness on $C_l^{\rm{kSZ}}$, higher minimum halo mass cases develop increased patchiness in IGM which in turn increases the kSZ angular power spectra.}
\label{tau54}
\end{figure*}

\begin{figure*}
\centering
{\includegraphics[trim={0cm 0cm 0cm 0cm},clip,width=1.0\textwidth]{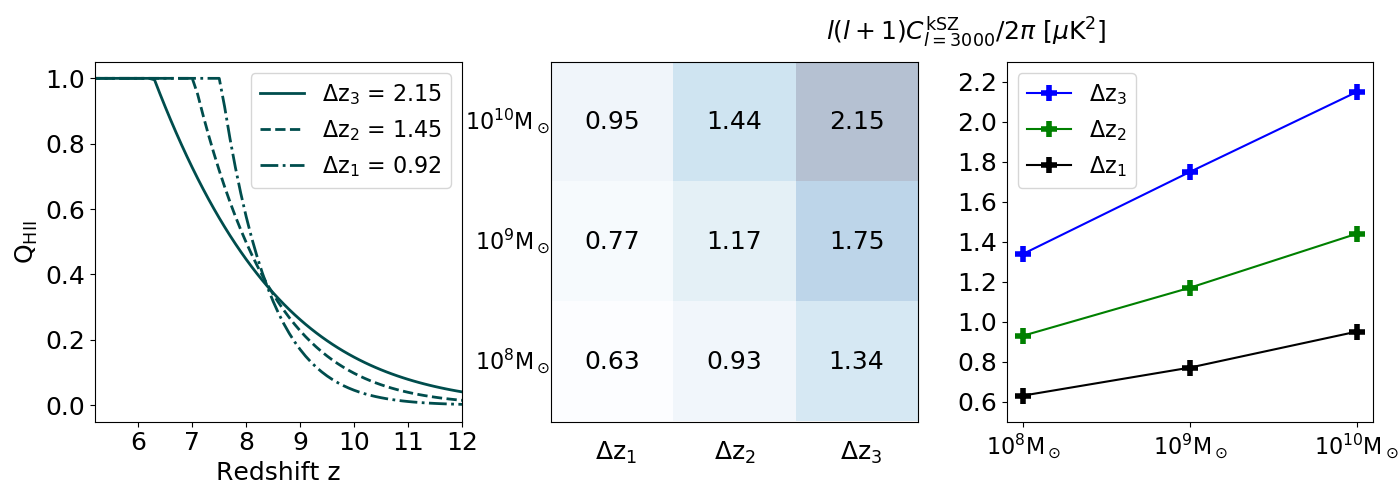}}
\caption{Same as Fig \ref{tau54} but for $\tau = 0.061$.}
\label{tau61}
\end{figure*}

\begin{figure*}
\centering
{\includegraphics[trim={0cm 0cm 0cm 0cm},clip,width=0.8\textwidth]{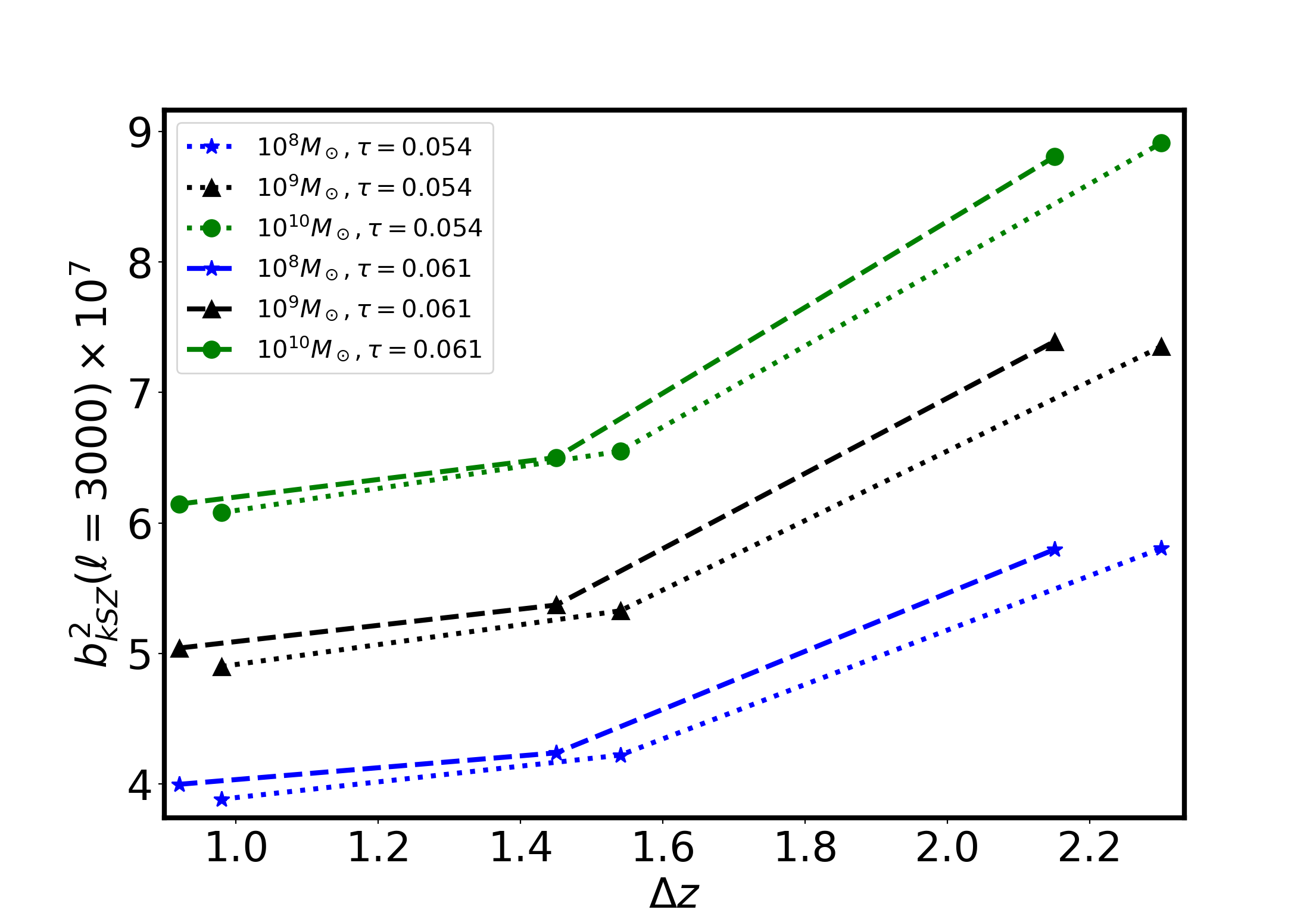}}
\caption{We show the values of the $b^2_{\rm kSZ} (l=3000)$, see equation (\ref{bkSZ}), as  function of $\Delta z$ for different models of reionization considered in this analysis. The bias parameter exhibits a stronger dependence on the minimum halo mass than the on the width of reionization $\Delta z$ and mean value of optical depth $\tau$.}
\label{biasksz}
\end{figure*}

The kSZ contribution arising from patchy ionization fields during the EoR to the CMB angular power spectrum is shown in Figure~\ref{kSZ_Cl} for the three cases M8, M9 and M10, respectively using the Limber's approximation (equation~\ref{Cl_eq}). We also show the signal arising from a homogeneous reionization where the reionization history is identical to the fiducial one, but the ionization fraction throughout the IGM is assumed to be homogeneous. Clearly, the patchiness enhances the amplitude of $C_l$ by over an order of magnitude at the angular scales relevant for the kSZ probes \citep{Park_2013}. Also, as anticipated from the discussions earlier in the paper, the signal amplitude is larger for higher values of $M_{\rm min}$.

It has been often suggested that \citep{Mcquinn_2005, Zahn_2012, Battaglia_2013} the amplitude of the kSZ signal is dictated primarily by two parameters: the redshift of half reionization ($z_{50\%}$) and the duration of reionization ($\Delta z = z_{75\%} - z_{25\%}$).  \citet{Battaglia_2013} calculated the kSZ power spectrum for $z>5.5$ and obtained the following scaling relation:
\begin{eqnarray}
D_{l=3000}^{\rm{kSZ}} &\equiv& \frac{l(l+1)}{2\pi}C_{l=3000}^{\rm{kSZ}} 
\nonumber \\
&\approx& 2.02 \mu{\rm K}^2 \left[ \left( \frac{1+\bar{z}}{11}\right) -0.12 \right]\left( \frac{\Delta z}{1.05}\right)^{0.47},
\label{battaglia}
\end{eqnarray} 
where $\bar{z}$ is the mean value for the reionization-redshift field, which is approximately equal to the redshift of half reionization $z_{50\%}$. For our fiducial reionization history, the above scaling relation gives $D_{l=3000}^{\rm{kSZ}} = 1.78 \mu \rm{K}^2$. The values we obtain for the models M8, M9 and M10 are $D_{l=3000}^{\rm{kSZ}} = 1.38 \mu \rm{K}^2$, $1.75 \mu \rm{K}^2$ and $2.19 \mu \rm{K}^2$, respectively. Since the three models have the identical reionization history, it is clear that the kSZ signal \emph{cannot} be described simply by two parameters $\bar{z}$ and $\Delta z$. Rather one needs to additionally include the information on the patchiness of the ionized regions.

To understand this further, we consider three reionization histories with the same $\tau =  0.054$ but having different $\Delta z$. These histories are shown in the left-hand panel of Figure~\ref{tau54}. Note that the model with $\Delta z = \Delta z_3 = 2.3$ corresponds to the fiducial reionization history, while the other two models have more rapid reionization. For each of these histories, we consider three values of $M_{\rm min}$, thus giving us a set of nine models. The values of $D_{l=3000}^{\rm{kSZ}}$ for these models are shown in the middle panel of Figure~\ref{tau54}. The duration of reionization increases the amplitude of $D_{l=3000}^{\rm{kSZ}}$, which is in qualitative agreement with the scaling relation \citet{Battaglia_2013}.\footnote{Note that we compute the signal by setting the lower limit of the integral in equation (\ref{Cl_eq}) to the redshift corresponding to the end of reionization so that the post-reionization OV component is not included. In contrast, the scaling relation of \citet{Battaglia_2013} is derived by setting the same lower limit of $z = 5.5$.} Additionally, the patchiness introduces another dimension in the kSZ anisotropy where increased patchiness (i.e., increased $M_{\rm min}$) increases the $D_{l=3000}^{\rm{kSZ}}$ value. It establishes the fact that a simple scaling relation based on half~-reionization redshift $\bar{z}$ and duration $\Delta z$ is not adequate to determine the kSZ power when dealing with patchy reionization. Our prediction estimates $\sim 55-60\%$ increase in the kSZ power at $l=3000$ when the $M_{\rm min}$ is hiked from $10^8 {\rm M}_{\odot}$ to $10^{10} {\rm M}_{\odot}$ (right panel of Figure \ref{tau54}). 

To see the effect of the CMB optical depth $\tau$ on the signal, we carry out an identical exercise for the value $\tau = 0.061$ which is the $1-\sigma$ upper limit provided by \citet{Planck:2018}. As before, we estimate the kSZ power for three reionization scenarios of varying duration (left panel of Figure~\ref{tau61}). For each of them, we hike the degree of patchiness from M8 to M10. The $D_{l=3000}^{\rm{kSZ}}$ estimates for this case are shown in the middle panel of Figure~\ref{tau61}. In this case, the shortest reionization scenario $\Delta z_1$ results in $\sim 50$\% increase in $D_{l=3000}^{\rm{kSZ}}$ from M8 to M10; whereas for $\Delta z_3$, the increment is $\sim 60\%$ (the right panel of Figure~\ref{tau61}).

In summary, the results from simulations show that spatial fluctuations in the electron density during the EoR, and hence the nature of the reionization sources, play a crucial role in determining the kSZ signal, which is not captured by the scaling relation mentioned in equation (\ref{battaglia}). It is thus important to derive a more detailed scaling relation which can account for this additional parameter. 

To do this, we first characterize the fluctuations in the electron density in terms of a bias. Let us define the bias of the ${\bf q}_{\perp}$ field as
\begin{equation}
b^2_{q_{\perp}}(k, \chi(z)) \equiv \frac{P_{q_\perp}(k, \chi (z))}{P_{\rm DM}(k, \chi(z))},
\end{equation}
which is a function of both $k$ and $z$. The main advantage of using the bias is that it normalizes the electron density fluctuations with respect to DM power spectrum and thus helps in mitigating the dependence on overall amplitude of matter power spectrum. The quantity relevant for the kSZ signal is the ``mean kSZ bias'' at a angular multipole $l$ defined as
\begin{eqnarray}
b^2_{\rm kSZ}(l) &\equiv& \frac{\int_{z_{\rm end}}^{z_{\rm beg}} dz~b^2_{q_{\perp}}(k = l / \chi(z), \chi(z))}{\int_{z_{\rm end}}^{z_{\rm beg}} dz}
\nonumber \\
&=& \frac{1}{z_{\rm beg} - z_{\rm end}} \int_{z_{\rm end}}^{z_{\rm beg}} dz~\frac{P_{q_\perp}(k=l / \chi(z), \chi (z))}{P_{\rm DM}(k=l / \chi(z), \chi(z))},
\nonumber \\
\label{bkSZ}
\end{eqnarray}
where the integration limits on $z$ correspond to the beginning and end of reionization. In reality, the above integral is calculated by summing over redshift bins where $N$-body simulation snapshots are available. The dependence of $b^2_{\rm kSZ}(l = 3000)$ on different parameters of reionization is shown in Figure \ref{biasksz}. As expected, the bias increases with increasing $\Delta z$ and with increasing $M_{\rm min}$. There is also a mild dependence on $\tau$ where the bias increases with increasing $\tau$.

\begin{figure*}
\centering
{\includegraphics[trim={0cm 0cm 0cm 0cm},clip,width=0.7\textwidth]{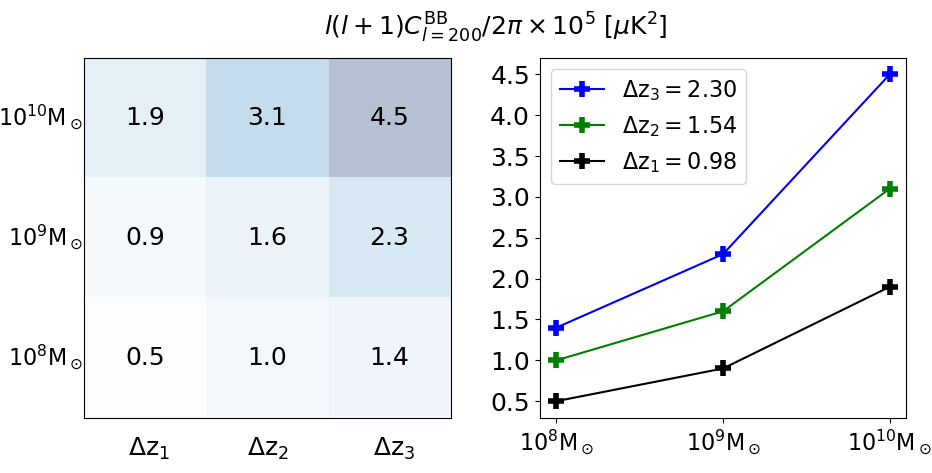}}
\caption{Dependence of $B$-mode polarization angular power spectra (calculated using the Limber's approximation mentioned in equation~\ref{bb-power-limber}) on reionization parameters. The reionization models considered are the same as those in Figure~\ref{tau54}. Left panel: : 2d distribution of $l(l+1)C_{l=200}^{\rm{BB}}/2\pi$ on various minimum halo mass and $\Delta z$ cases. Right panel: Effect of patchiness on $C_l^{\rm{BB}}$, higher minimum halo mass cases develop increased patchiness in IGM which in turn increases the amplitude of B-mode angular power spectra.}
\label{tau54_ClBB}
\end{figure*}

Using the results from our simulations, we obtain a new scaling relation for the kSZ amplitude
\begin{eqnarray}
\label{newscaling}
D^{\rm kSZ}_{l=3000} &\approx & 0.65 \mu {\rm K}^2\bigg(\frac{0.097+ \tau}{0.151}\bigg)\bigg(\frac{\Delta z}{1.0}\bigg)^{0.54}\nonumber\\ 
&&\times \bigg(\frac{b^2_{\rm kSZ}(l=3000)}{4.0 \times 10^{-7}}\bigg)^{0.92},
\end{eqnarray}
expressed in terms of the mean optical depth $\bar \tau$, duration of the EoR $\Delta z$ and the mean bias $b^2_{\rm kSZ} (l=3000)$. Another novelty of this scaling relation is that it is expressed in terms of $\tau$, which can be measured accurately from the reionization bump of CMB $E$-mode polarization signal. We have elaborated the procedure for obtaining the scaling relation in Appendix \ref{scal}. The above scaling relation shows a maximum of $13\%$ deviation in the amplitude of the kSZ signal in comparison to the results from simulations (shown in Figure~\ref{tau54_resBB} in the appendix \ref{scal}). This new scaling relation indicates that an observed $D^{\rm kSZ}_{l=3000}$ cannot be related only with the $\Delta z$ parameter and mean redshift of reionization, but it also is quite sensitive to the amplitude of fluctuations in the electron density denoted by the term $b^2_{\rm kSZ}$ (as can be seen from the power-law exponents of the respective terms).

\begin{figure*}
\centering
{{\includegraphics[trim={0cm 0cm 0cm 0cm},clip,width=0.7\textwidth]{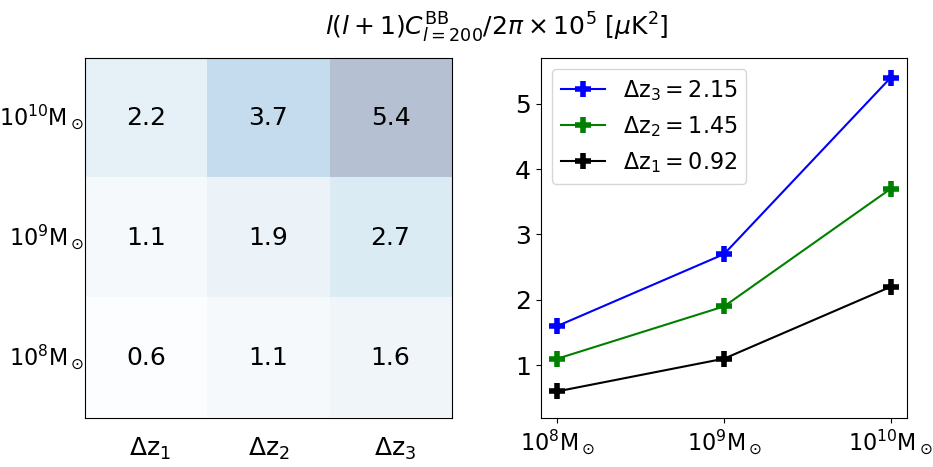}}}
\caption{Same as Figure~\ref{tau54_ClBB} but for $\tau = 0.061$.}
\label{tau61_ClBB}
\end{figure*}

 In order to break the degeneracy between the width of reionization and spatial fluctuations in electron density, we need to combine additional dataset as well as by exploring higher order correlation functions \citep{Smith:2016lnt, Ferraro:2018izc}. This new scaling relation provided in this paper will be useful for the interpretation of CMB data from the high resolution CMB experiments such as  AdvACTPol \citep{2016ApJS..227...21T}, SPT-3G \citep{Benson:2014qhw}, Simons Observatory \citep{Ade:2018sbj}, and CMB-S4 \citep{Abazajian:2019eic}. A recent measurement of kSZ signal during reionization and the corresponding bounds on the width of reionization $\Delta z$ \citep{2020arXiv200206197R} is subject to vary due to the inclusion of the contribution from patchiness in electron density during reionization. However, it is crucial to correctly model the OV part before applying the scaling relation to the data \citep[see, e.g.,][]{2020arXiv200206197R}.
 In our recent work  \citep{Choudhury:2020kzh}, by using a physical model of reionization and scaling laws for the OV part according to \citet{2012ApJ...756...15S}, tight constraints on $\Delta z$ and kSZ bias parameter (mentioned in Eq. \ref{newscaling}) are obtained.
 In a future work, we will develop estimators to measure both spatial fluctuations and reionization from CMB data.


\subsection{Impact on $B$-mode polarization power spectrum}
The nature of secondary anisotropies in the $B$-mode polarization during the EoR at $l=200$ is shown in Figure~\ref{tau54_ClBB}
and \ref{tau61_ClBB} for two cases of the optical depth $\tau =0.54$ and $\tau= 0.61$ respectively. The $B$-mode angular power spectra are estimated with the Limber's approximation mentioned in equation~\ref{bb-power-limber}. Each case is obtained for nine different reionization scenarios by varying the parameters such as $\Delta z$ and the minimum halo masses. These sets of total $18$ different cases make it possible to show the dependence of the amplitude of the secondary $B$-mode polarization signal originating from patchy reionization. Our results show that fluctuations in the $B$-mode polarization signal can get stronger with the increase in the width of the reionization $\Delta z$ and for the increase in the spatial fluctuations of the electron density during the EoR. The impact of inhomogeneities in electron density leads to a stronger impact in the amplitude of the $B$-mode polarization signal. For the limited cases of reionization considered in this analysis, we find that the amplitude of the $B$-mode polarization signal can increase up to nearly nine times from the case of small $\Delta z$ and minimum spatial fluctuations in electron density (M8 case) to large $\Delta z$ and maximum spatial fluctuations in electron density (M10 case).

\begin{figure*}
\centering
{\includegraphics[trim={0cm 0cm 0cm 0cm},clip,width=0.8\textwidth]{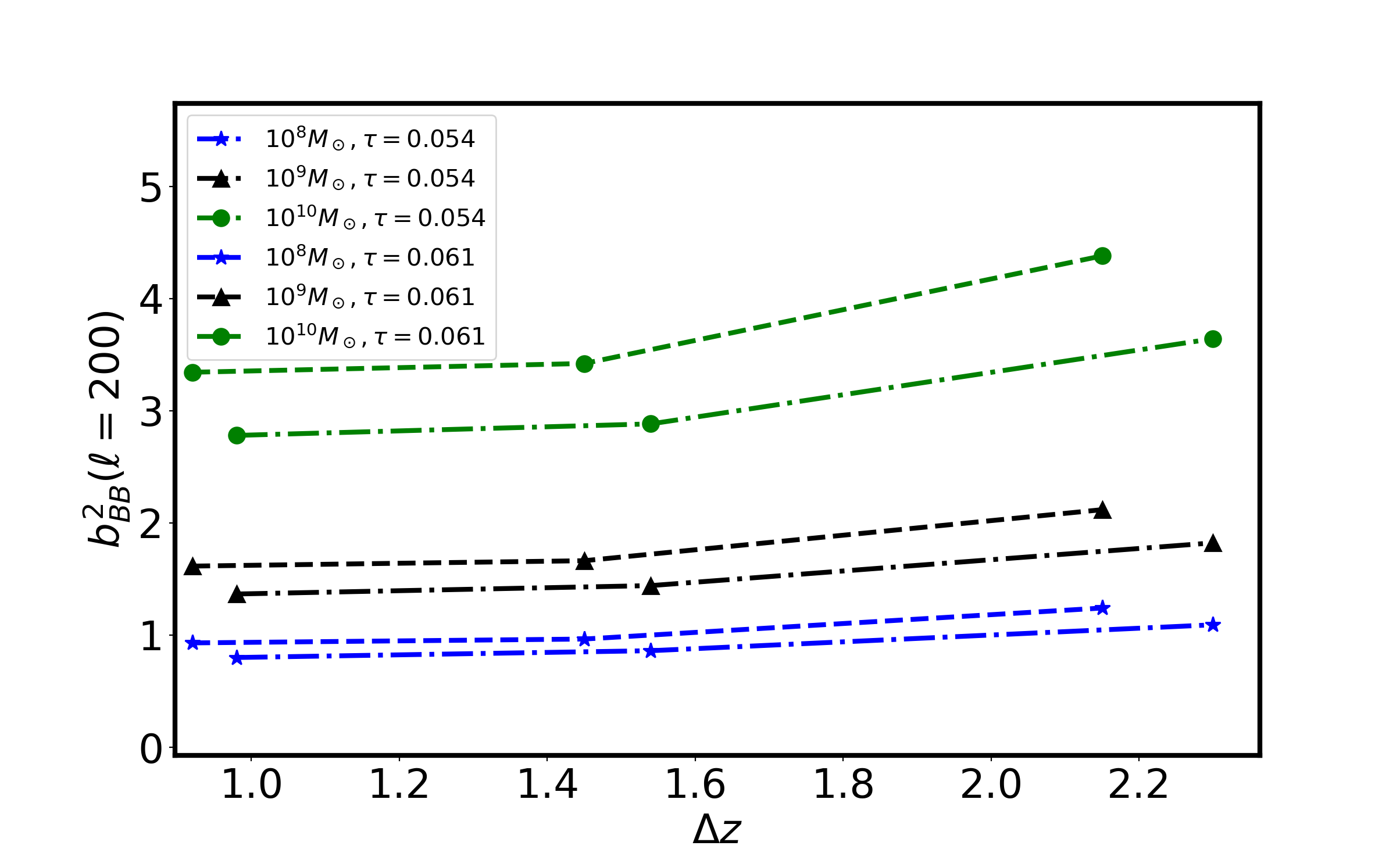}}
\caption{We show the values of the $b^2_{BB} (l=200)$, see equation (\ref{bBB}), as  function of $\Delta z$ for different models of reionization considered in this analysis. Similar to the behavior for kSZ, the bias parameter shows a stronger dependence on the minimum halo mass than the on the width of reionization $\Delta z$ and mean value of optical depth $\tau$.}
\label{biasbb}
\end{figure*}

As we did in the case of kSZ signal, let us define the bias in the electron density as
\begin{equation}
b^2_{ee}(k, \chi(z)) \equiv \frac{P_{ee}(k, \chi (z))}{P_{\rm DM}(k, \chi(z))},
\end{equation}
which can then be used to define the ``mean $BB$ bias'' at a angular multipole $l$
\begin{eqnarray}
b^2_{BB}(l) &\equiv& \frac{\int_{z_{\rm end}}^{z_{\rm beg}} dz~b^2_{ee}\left(k = \frac{l + 1/2}{\chi(z)}, \chi(z)\right)}{\int_{z_{\rm end}}^{z_{\rm beg}} dz}
\nonumber \\
&=& \frac{1}{z_{\rm beg} - z_{\rm end}} \int_{z_{\rm end}}^{z_{\rm beg}} dz~\frac{P_{ee}\left(k = \frac{l + 1/2}{\chi(z)}, \chi(z)\right)}{P_{\rm DM}\left(k = \frac{l + 1/2}{\chi(z)}, \chi(z)\right)}.
\nonumber \\
\label{bBB}
\end{eqnarray}
The range of values of $b^{2}_{BB}(l=200)$ are shown in Figure~\ref{biasbb} indicating a maximum variation up to a factor of four for the simulation cases considered in this analysis. In particular, we see a clear dependence of the bias on the value of $M_{\rm min}$. 

Using the set of simulations, we have obtained a scaling relation for the amplitude of the $B$-mode polarization  signal ($D^{BB}_l\equiv l(l+1)C^{BB}_l/2\pi$): 
\begin{equation}\label{newscalingbmode}
D^{BB}_{l=200}\approx 6.6\, {\rm n K}^2 \bigg(\frac{0.15+ \tau}{0.204}\bigg)\bigg(\frac{\Delta z}{0.98}\bigg)^{0.78}\bigg(\frac{b^2_{\rm BB}(l=200)}{0.93}\bigg)^{0.99}.
\end{equation}
This new scaling relation will be useful to estimate contribution of patchy reionization from the $B$-mode polarization data of the upcoming missions \citep{Ade:2018sbj, 2018JLTP..193.1048S, Abazajian:2019eic}. The procedure followed to obtained the scaling relation is mentioned in Appendix \ref{scal}. The scaling relation fits well with the simulation results and shows a maximum of $\sim 14\%$ departure from the results obtained using simulations (shown in Figure~\ref{tau54_resBB}).

Similar to kSZ, the $B$-mode polarization signal is also affected by both $\Delta z$ and spatial fluctuations in electron density during the EoR. However, the dependence of kSZ and $B$-mode polarization on $\Delta z$ and spatial fluctuations in electron density are different [compare the power-law exponents in equations (\ref{newscaling}) and (\ref{newscalingbmode})]. As a result, by combining both kSZ and the $B$-mode polarization signal, we can learn about $\Delta z$ and spatial fluctuations in electron density. In a future work, we will explore the estimators by combining kSZ and $B$-mode polarization signal to understand the EoR.

Additional fluctuations due to patchy reionization can also bias the inferred value of the amplitude of primordial gravitational waves denoted by $r$ (tensor to scalar ratio) \citep{Mortonson:2007, Mukherjee_2019}. In our previous study from simulations \citep{Mukherjee_2019}, we have shown that the bias in the value of $r$ gets stronger for reionization driven by massive halos and can be comparable with the error bar of the upcoming CMB missions \citep[see also][]{Roy:2020cqn}. A joint estimation of the kSZ signal, $E$-mode polarization and $B$-mode polarization signal will also be useful to mitigate the contamination in the primordial $B$-mode polarization signal.

\subsection{21cm angular power spectrum from simulations}

\begin{figure*}
\centering
{\includegraphics[trim={4cm 4.7cm 3.9cm 4.1cm},clip,width=1.\textwidth]{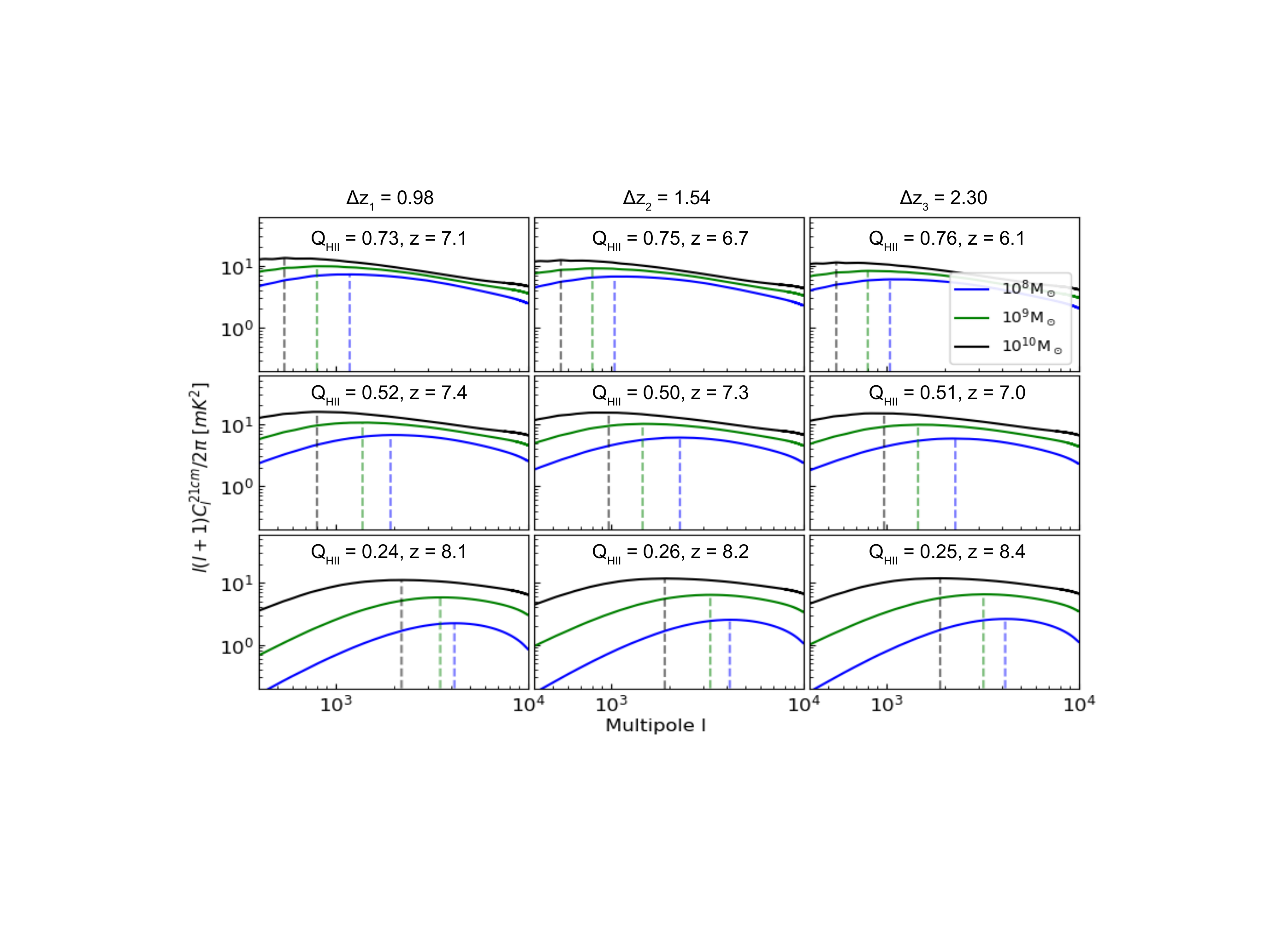}}
\caption{The 21~cm angular power spectra under flat-sky approximation derived from the 3D power spectrum of HI brightness fluctuation $P_{\rm{HI}}(k)$ at various redshifts for the three reionization histories considered in this paper. The location of peak in each curve indicates the characteristic bubble size. With decreasing redshift (bottom to top), the peaks shift to lower $l$ which implies increase in the size of HII regions as reionization progresses. At a fixed redshift, peaks for $10^9$ M$_\odot$ and $10^{10}$ M$_\odot$ occur at a lower CMB multipoles $l$ compared to the $10^8$ M$_\odot$ which is also indicative of larger bubble size for higher minimum halo mass cases.}
\label{21cm_Cl}
\end{figure*}

So far, we have been concentrating on the CMB signals which are essentially integrated along the light cone. Hence it is often not straightforward to connect the angular scales or multipoles $l$ to physical quantities like the characteristic ionized bubble size. As a complementary probe, we consider the 21~cm probe of HI, which has the advantage that one can study the patchy EoR signal arising from individual cosmic epochs. This would allow us to understand the $l$-values which are sensitive to the bubble sizes.

In order to compare the 21~cm signal with the CMB probes, we concentrate on the angular power spectrum $C_l^{\rm{21cm}}(z) $defined, under flat-sky approximation, as \citep{Bharadwaj_2004, Dutta_2007}:
\begin{equation}
   C_l^{\rm{21cm}}(z) = \frac{1}{\pi \chi^2} \int dk_\parallel P_{\delta T_b}({\bf k},z),
   \label{Cl21}
\end{equation}
where $P_{\delta T_b}({\bf k},z)$ is the power spectrum of the $\delta T_b$ field defined in equation (\ref{21cm}). The vector {\bf k} has components along the line of sight and on the plane of the sky as $k_\parallel$ and $l/\chi$ respectively. 

Shown in Figure~\ref{21cm_Cl} are the $C_l^{\rm{21cm}}(z)$ vs $l$ plot for various cases considered in this paper. The M9 and M10 cases exhibit higher power compared to the M8 case. Also, the locations of the peak suggest that both those cases manifest larger characteristic ionized bubble sizes as they occur at lower $l$ values. As reionization proceeds, the bubbles grow up in size, which is evident as the locations of the peak shift at a lower $l$ value with decreasing redshift. Thus we conclude that the minimum halo mass parameter affects the characteristic bubble size. Also, the increase in the $C_l^{\rm{21cm}}$ with higher $M_{\rm{min}}$ further confirms that it has an amplifying effect on the patchiness in the IGM (also supported by Figure~\ref{21cm_cones}). Interestingly, the $l$-values corresponding to the typical bubble sizes during the EoR correspond to the scales probes by the kSZ experiments, hence cross-correlating the two signals should reveal further details of reionization \citep{2005MNRAS.360.1063S} which we would take up in a future project. A crucial advantage of such cross-correlation studies is that one can minimize the systematics arising from individual experiments. However, a straightforward kSZ-21cm cross-correlation can result in a cancellation effect as the ionized regions are equally likely to move towards or away along the line of sight. To avoid such cancellation, hybrid approaches such as kSZ$^2$-21cm \citep{Ma_2018} or higher-order correlation studies will be explored in a future work.

\section{Conclusion}\label{Conclusion}
Cosmic reionization is a crucial phenomenon finished by redshift $z \approx 5.5$ \citep{BarkanaLoeb:2001,BarkanaLoeb:2004,Furlanetto_2004,Wyithe_2003, 2018PhR...780....1D}. A detailed understanding of the EoR is crucial to understand several astrophysical and cosmological processes. In this paper, we explore the impact of patchy reionization on cosmological observables such as kSZ temperature anisotropy, $B$-mode polarization signal, and 21~cm brightness temperature fluctuation using a photon-conserving semi-numerical simulation called \texttt{SCRIPT} \citep{Choudhury:2018}. We parametrize the EoR in terms of three parameters, the optical depth $\tau$, duration of the reionization $\Delta z$ and spatial fluctuations in the distribution of electron density during the EoR. From a set of $18$ semi-numerical simulations by varying $\tau$, $\Delta z$ and minimum halo mass driving the reionization, we estimate the angular power signal spectrum of kSZ, $B$-mode polarization, and 21-cm signal. 

We show that the spatial fluctuations in the electron density during cosmic reionization (which we call as patchiness) is going to play a crucial role in determining the amplitude of angular power spectrum of kSZ, $B$-mode polarization, and 21-cm signal. The amplitude of the kSZ power spectrum is usually considered to depend only on the width of the epoch of reionization $\Delta z$, and redshift of half-reionized Universe $z_{50\%}$ \citep[as shown in equation (\ref{battaglia}), taken from][]{Battaglia_2013}. However, we find that the patchiness creates an additional signal in the kSZ power spectrum and hence the observed amplitude of the kSZ signal cannot be related only to the $\Delta z$ and $z_{50\%}$ parameters. We have obtained a new scaling relation for the kSZ amplitude mentioned in equation (\ref{newscaling}), which captures the effect of spatial fluctuation and differs from the previously used relation mentioned in equation (\ref{battaglia}). A recent work by \citet{Gorce:2020pcy} have also pointed out the difference in the kSZ amplitude in comparison to the results by \citet{Battaglia_2013}. Results obtained by \citet{Gorce:2020pcy} are in agreement with the simulation method which we have used in this analysis.\footnote{In Figure 8 of \citet{Gorce:2020pcy}, they have shown that the kSZ fluctuations are more for $10^{8}$ $M_\odot$ than for the $10^{10}$ $M_\odot$ case. However, both these results are obtained for different reionization histories and are the possible reason for this counter-intuitive outcome.}

The width of the EoR $\Delta z$ and the patchiness during the epoch of cosmic reionization also leads to an observable effect in the $B$-mode polarization. The amplitude of the secondary $B$-mode polarization during the EoR increases with the increase in the value of $\Delta z$, and spatial fluctuations. The dependence of the amplitude of the $B$-mode polarization signal is stronger with an increase in the amplitude of fluctuations in the electron density than $\Delta z$ as shown by the scaling relation equation (\ref{newscalingbmode}). This new scaling relation will be useful to infer the contribution from patchy reionization from CMB B-mode polarization data of the upcoming missions \citep{Ade:2018sbj, 2018JLTP..193.1048S, Abazajian:2019eic}. 

The scaling relations we derived in the paper indicate two important aspects: (i) the measurement of kSZ signal and a value of $\tau$ is not sufficient to infer the value of $\Delta z$ from observations. It is crucial to also know the contribution from spatial fluctuations in electron density during the EoR. (ii) The contribution to the $B$-mode polarization depends on the reionization history and shows a variation up to a factor of nine in the amplitude of the signal (see Sec. \ref{Results}). The combination of both kSZ and $B$-mode polarization signal can help in breaking the degeneracy the $\Delta z$ and spatial fluctuations, as they follow different scaling relations. By combining the measurement from CMB-only measurements such as $E$-mode polarization (the reionization bump), $B$-mode polarization, and kSZ signal from the temperature field, we can make an accurate measurement of the optical depth $\tau$, along with $\Delta z$ and patchiness during the EoR. With the availability of the data of 21 cm signal from different redshifts, the information on spatial fluctuations in the electron density can be improved and can be related to the source properties. The cross-correlation study between 21 cm and the kSZ signal is also going to be a useful avenue to study the fluctuations in the electron density. 

In summary, we would like to point out that using only the upcoming CMB observables, secondary anisotropies generated during the epoch of cosmic reionization can be explored. Not considering patchiness in the electron density during the EoR in the analysis of CMB data can lead to an inaccurate inference of the duration of reionization. In a future analysis, we will address the measurability of the reionization related parameters from the joint analysis of high angular resolution ground-based CMB experiments \citep{2016ApJS..227...21T, Benson:2014qhw, Ade:2018sbj, Abazajian:2019eic} and space-based CMB experiments \citep{2018JLTP..193.1048S}.

\section*{Acknowledgement}
We thank the anonymous referee for the helpful comments on the draft that have contributed to improving this paper. SP acknowledges SARAO for support through the SKA postdoctoral fellowship.
SM would like to thank Francois Bouchet, Masashi Hazumi, Eiichiro Komatsu, Marius Millea, Joseph Silk, and Kimmy Wu for useful conversations. SM would also like to thank Simone Ferraro for pointing to useful references. 
 A part of the computational works are carried out at the Horizon cluster hosted by Institut d'Astrophysique de Paris. We thank Stephane Rouberol for smoothly running the Horizon cluster.
The work of SM is supported by the Labex ILP (reference ANR-10-LABX-63) part of the Idex SUPER,  received financial state aid managed by the Agence Nationale de la Recherche, as part of the programme Investissements d'avenir under the reference ANR-11-IDEX-0004-02, and also by the research program Innovational Research Incentives Scheme (Vernieuwingsimpuls), which is financed by the Netherlands Organization for Scientific Research through the NWO VIDI Grant No. 639.042.612-Nissanke. 
TRC acknowledges support of the Department of Atomic Energy, Government of India, under project no. 12-R\&D-TFR-5.02-0700.
In this analysis, we have used the  following packages: IPython \citep{PER-GRA:2007}, Matplotlib \citep{Hunter:2007},  NumPy \citep{2011CSE....13b..22V}, and SciPy \citep{scipy}. 

\section*{Data availability}
The data underlying this article will be shared on reasonable request to the corresponding author (SP).

\bibliography{references}

\begin{thebibliography}{}
\makeatletter
\relax
\def\mn@urlcharsother{\let\do\@makeother \do\$\do\&\do\#\do\^\do\_\do\%\do\~}
\def\mn@doi{\begingroup\mn@urlcharsother \@ifnextchar [ {\mn@doi@}
  {\mn@doi@[]}}
\def\mn@doi@[#1]#2{\def\@tempa{#1}\ifx\@tempa\@empty \href
  {http://dx.doi.org/#2} {doi:#2}\else \href {http://dx.doi.org/#2} {#1}\fi
  \endgroup}
\def\mn@eprint#1#2{\mn@eprint@#1:#2::\@nil}
\def\mn@eprint@arXiv#1{\href {http://arxiv.org/abs/#1} {{\tt arXiv:#1}}}
\def\mn@eprint@dblp#1{\href {http://dblp.uni-trier.de/rec/bibtex/#1.xml}
  {dblp:#1}}
\def\mn@eprint@#1:#2:#3:#4\@nil{\def\@tempa {#1}\def\@tempb {#2}\def\@tempc
  {#3}\ifx \@tempc \@empty \let \@tempc \@tempb \let \@tempb \@tempa \fi \ifx
  \@tempb \@empty \def\@tempb {arXiv}\fi \@ifundefined
  {mn@eprint@\@tempb}{\@tempb:\@tempc}{\expandafter \expandafter \csname
  mn@eprint@\@tempb\endcsname \expandafter{\@tempc}}}

\bibitem[\protect\citeauthoryear{Abazajian et~al.}{Abazajian
  et~al.}{2019}]{Abazajian:2019eic}
Abazajian K.,  et~al., 2019, arXiv:1907.04473

\bibitem[\protect\citeauthoryear{Aghanim, Majumdar  \& Silk}{Aghanim
  et~al.}{2008}]{Aghanim_2008}
Aghanim N.,  Majumdar S.,   Silk J.,  2008, \mn@doi [Reports on Progress in
  Physics] {10.1088/0034-4885/71/6/066902}, 71, 066902

\bibitem[\protect\citeauthoryear{Aguirre et~al.}{Aguirre
  et~al.}{2018}]{Ade:2018sbj}
Aguirre J.,  et~al., 2018, arXiv:1808.07445

\bibitem[\protect\citeauthoryear{Alvarez}{Alvarez}{2016}]{Alvarez_2016}
Alvarez M.~A.,  2016, \mn@doi [The Astrophysical Journal]
  {10.3847/0004-637x/824/2/118}, 824, 118

\bibitem[\protect\citeauthoryear{{Barkana} \& {Loeb}}{{Barkana} \&
  {Loeb}}{2001}]{BarkanaLoeb:2001}
{Barkana} R.,  {Loeb} A.,  2001, \mn@doi [\physrep]
  {10.1016/S0370-1573(01)00019-9}, \href
  {http://adsabs.harvard.edu/abs/2001PhR...349..125B} {349, 125}

\bibitem[\protect\citeauthoryear{{Barkana} \& {Loeb}}{{Barkana} \&
  {Loeb}}{2004}]{BarkanaLoeb:2004}
{Barkana} R.,  {Loeb} A.,  2004, \apj, \href
  {http://stacks.iop.org/0004-637X/609/i=2/a=474} {609, 474}

\bibitem[\protect\citeauthoryear{{Barkana} \& {Loeb}}{{Barkana} \&
  {Loeb}}{2006}]{BarkanaLoeb:2006}
{Barkana} R.,  {Loeb} A.,  2006, \mn@doi [\mnras]
  {10.1111/j.1745-3933.2006.00222.x}, \href
  {http://adsabs.harvard.edu/abs/2006MNRAS.372L..43B} {372, L43}

\bibitem[\protect\citeauthoryear{{Battaglia}, {Natarajan}, {Trac}, {Cen}  \&
  {Loeb}}{{Battaglia} et~al.}{2013}]{Battaglia_2013}
{Battaglia} N.,  {Natarajan} A.,  {Trac} H.,  {Cen} R.,   {Loeb} A.,  2013,
  \mn@doi [\apj] {10.1088/0004-637X/776/2/83}, \href
  {https://ui.adsabs.harvard.edu/abs/2013ApJ...776...83B} {776, 83}

\bibitem[\protect\citeauthoryear{Benson et~al.}{Benson
  et~al.}{2014}]{Benson:2014qhw}
Benson B.~A.,  et~al., 2014, \mn@doi [Proc. SPIE Int. Soc. Opt. Eng.]
  {10.1117/12.2057305}, 9153, 91531P

\bibitem[\protect\citeauthoryear{{Bharadwaj} \& {Ali}}{{Bharadwaj} \&
  {Ali}}{2004}]{Bharadwaj_2004}
{Bharadwaj} S.,  {Ali} S.~S.,  2004, \mn@doi [\mnras]
  {10.1111/j.1365-2966.2004.07907.x}, \href
  {https://ui.adsabs.harvard.edu/abs/2004MNRAS.352..142B} {352, 142}

\bibitem[\protect\citeauthoryear{Calabrese et~al.}{Calabrese
  et~al.}{2014}]{Calabrese:2014gwa}
Calabrese E.,  et~al., 2014, \mn@doi [JCAP] {10.1088/1475-7516/2014/08/010},
  08, 010

\bibitem[\protect\citeauthoryear{{Choudhury} \& {Paranjape}}{{Choudhury} \&
  {Paranjape}}{2018}]{Choudhury:2018}
{Choudhury} T.~R.,  {Paranjape} A.,  2018, \mn@doi [MNRAS]
  {10.1093/mnras/sty2551}, \href
  {https://ui.adsabs.harvard.edu/#abs/2018MNRAS.481.3821C} {481, 3821}

\bibitem[\protect\citeauthoryear{{Choudhury}, {Ferrara}  \&
  {Gallerani}}{{Choudhury} et~al.}{2008}]{2008MNRAS.385L..58C}
{Choudhury} T.~R.,  {Ferrara} A.,   {Gallerani} S.,  2008, \mn@doi [\mnras]
  {10.1111/j.1745-3933.2008.00433.x}, \href
  {https://ui.adsabs.harvard.edu/abs/2008MNRAS.385L..58C} {385, L58}

\bibitem[\protect\citeauthoryear{{Choudhury}, {Haehnelt}  \&
  {Regan}}{{Choudhury} et~al.}{2009}]{Choudhury:2009}
{Choudhury} T.~R.,  {Haehnelt} M.~G.,   {Regan} J.,  2009, \mn@doi [MNRAS]
  {10.1111/j.1365-2966.2008.14383.x}, \href
  {http://adsabs.harvard.edu/abs/2009MNRAS.394..960C} {394, 960}

\bibitem[\protect\citeauthoryear{{Choudhury}, {Datta}, {Majumdar}, {Ghara},
  {Paranjape}, {Mondal}, {Bharadwaj}  \& {Samui}}{{Choudhury}
  et~al.}{2016}]{2016JApA...37...29C}
{Choudhury} T.~R.,  {Datta} K.,  {Majumdar} S.,  {Ghara} R.,  {Paranjape} A.,
  {Mondal} R.,  {Bharadwaj} S.,   {Samui} S.,  2016, \mn@doi [Journal of
  Astrophysics and Astronomy] {10.1007/s12036-016-9403-z}, \href
  {https://ui.adsabs.harvard.edu/abs/2016JApA...37...29C} {37, 29}

\bibitem[\protect\citeauthoryear{Choudhury, Mukherjee  \& Paul}{Choudhury
  et~al.}{2020}]{Choudhury:2020kzh}
Choudhury T.~R.,  Mukherjee S.,   Paul S.,  2020

\bibitem[\protect\citeauthoryear{Crawford et~al.}{Crawford
  et~al.}{2014}]{Crawford:2013uka}
Crawford T.,  et~al., 2014, \mn@doi [Astrophys. J.]
  {10.1088/0004-637X/784/2/143}, 784, 143

\bibitem[\protect\citeauthoryear{{Datta}, {Choudhury}  \& {Bharadwaj}}{{Datta}
  et~al.}{2007}]{Dutta_2007}
{Datta} K.~K.,  {Choudhury} T.~R.,   {Bharadwaj} S.,  2007, \mn@doi [\mnras]
  {10.1111/j.1365-2966.2007.11747.x}, \href
  {https://ui.adsabs.harvard.edu/abs/2007MNRAS.378..119D} {378, 119}

\bibitem[\protect\citeauthoryear{{Datta}, {Mellema}, {Mao}, {Iliev}, {Shapiro}
  \& {Ahn}}{{Datta} et~al.}{2012}]{Datta:2012}
{Datta} K.~K.,  {Mellema} G.,  {Mao} Y.,  {Iliev} I.~T.,  {Shapiro} P.~R.,
  {Ahn} K.,  2012, \mn@doi [\mnras] {10.1111/j.1365-2966.2012.21293.x}, \href
  {http://adsabs.harvard.edu/abs/2012MNRAS.424.1877D} {424, 1877}

\bibitem[\protect\citeauthoryear{{Dayal} \& {Ferrara}}{{Dayal} \&
  {Ferrara}}{2018}]{2018PhR...780....1D}
{Dayal} P.,  {Ferrara} A.,  2018, \mn@doi [\physrep]
  {10.1016/j.physrep.2018.10.002}, \href
  {http://adsabs.harvard.edu/abs/2018PhR...780....1D} {780, 1}

\bibitem[\protect\citeauthoryear{DeBoer et~al.,}{DeBoer
  et~al.}{2017}]{DeBoer_2017}
DeBoer D.~R.,  et~al., 2017, \mn@doi [Publications of the Astronomical Society
  of the Pacific] {10.1088/1538-3873/129/974/045001}, 129, 045001

\bibitem[\protect\citeauthoryear{Delabrouille et~al.}{Delabrouille
  et~al.}{2019}]{Delabrouille:2019thj}
Delabrouille J.,  et~al., 2019

\bibitem[\protect\citeauthoryear{Dunkley et~al.}{Dunkley
  et~al.}{2013}]{Dunkley:2013vu}
Dunkley J.,  et~al., 2013, \mn@doi [JCAP] {10.1088/1475-7516/2013/07/025}, 07,
  025

\bibitem[\protect\citeauthoryear{Dvorkin \& Smith}{Dvorkin \&
  Smith}{2009}]{Dvorkin:2008tf}
Dvorkin C.,  Smith K.~M.,  2009, \mn@doi [Phys. Rev.]
  {10.1103/PhysRevD.79.043003}, D79, 043003

\bibitem[\protect\citeauthoryear{{Dvorkin}, {Hu}  \& {Smith}}{{Dvorkin}
  et~al.}{2009}]{Dvorkin:2009}
{Dvorkin} C.,  {Hu} W.,   {Smith} K.~M.,  2009, \mn@doi [\prd]
  {10.1103/PhysRevD.79.107302}, \href
  {http://adsabs.harvard.edu/abs/2009PhRvD..79j7302D} {79, 107302}

\bibitem[\protect\citeauthoryear{Ferraro \& Smith}{Ferraro \&
  Smith}{2018}]{Ferraro:2018izc}
Ferraro S.,  Smith K.~M.,  2018, \mn@doi [Phys. Rev.]
  {10.1103/PhysRevD.98.123519}, D98, 123519

\bibitem[\protect\citeauthoryear{{Field}}{{Field}}{1958}]{Field:1958}
{Field} G.~B.,  1958, \mn@doi [Proceedings of the IRE]
  {10.1109/JRPROC.1958.286741}, \href
  {http://adsabs.harvard.edu/abs/1958PIRE...46..240F} {46, 240}

\bibitem[\protect\citeauthoryear{{Field}}{{Field}}{1959}]{Field:1959}
{Field} G.~B.,  1959, \mn@doi [\apj] {10.1086/146653}, \href
  {http://adsabs.harvard.edu/abs/1959ApJ...129..536F} {129, 536}

\bibitem[\protect\citeauthoryear{Furlanetto, Zaldarriaga  \&
  Hernquist}{Furlanetto et~al.}{2004}]{Furlanetto_2004}
Furlanetto S.~R.,  Zaldarriaga M.,   Hernquist L.,  2004, \mn@doi [The
  Astrophysical Journal] {10.1086/423028}, 613, 16

\bibitem[\protect\citeauthoryear{Geil \& Wyithe}{Geil \&
  Wyithe}{2008}]{Geil:2007rj}
Geil P.~M.,  Wyithe S.,  2008, \mn@doi [Mon. Not. Roy. Astron. Soc.]
  {10.1111/j.1365-2966.2008.13159.x}, 386, 1683

\bibitem[\protect\citeauthoryear{{Ghara}, {Datta}  \& {Choudhury}}{{Ghara}
  et~al.}{2015}]{Ghara:2015}
{Ghara} R.,  {Datta} K.~K.,   {Choudhury} T.~R.,  2015, \mn@doi [\mnras]
  {10.1093/mnras/stv1855}, \href
  {http://adsabs.harvard.edu/abs/2015MNRAS.453.3143G} {453, 3143}

\bibitem[\protect\citeauthoryear{{Gluscevic}, {Kamionkowski}  \&
  {Hanson}}{{Gluscevic} et~al.}{2013}]{2013PhRvD..87d7303G}
{Gluscevic} V.,  {Kamionkowski} M.,   {Hanson} D.,  2013, \mn@doi [\prd]
  {10.1103/PhysRevD.87.047303}, \href
  {https://ui.adsabs.harvard.edu/abs/2013PhRvD..87d7303G} {87, 047303}

\bibitem[\protect\citeauthoryear{Gorce, Ilić, Douspis, Aubert  \&
  Langer}{Gorce et~al.}{2020}]{Gorce:2020pcy}
Gorce A.,  Ilić S.,  Douspis M.,  Aubert D.,   Langer M.,  2020, arXiv:
  2004.06616

\bibitem[\protect\citeauthoryear{{Gruzinov} \& {Hu}}{{Gruzinov} \&
  {Hu}}{1998}]{1998ApJ...508..435G}
{Gruzinov} A.,  {Hu} W.,  1998, \mn@doi [\apj] {10.1086/306432}, \href
  {https://ui.adsabs.harvard.edu/abs/1998ApJ...508..435G} {508, 435}

\bibitem[\protect\citeauthoryear{Hanany et~al.}{Hanany
  et~al.}{2019}]{Hanany:2019lle}
Hanany S.,  et~al., 2019

\bibitem[\protect\citeauthoryear{{Howlett}, {Lewis}, {Hall}  \&
  {Challinor}}{{Howlett} et~al.}{2012}]{2012JCAP...04..027H}
{Howlett} C.,  {Lewis} A.,  {Hall} A.,   {Challinor} A.,  2012, \mn@doi [\jcap]
  {10.1088/1475-7516/2012/04/027}, \href
  {https://ui.adsabs.harvard.edu/abs/2012JCAP...04..027H} {2012, 027}

\bibitem[\protect\citeauthoryear{Hu}{Hu}{2000}]{Hu:2000}
Hu W.,  2000, \apj, 529, 12

\bibitem[\protect\citeauthoryear{Hu}{Hu}{2001}]{Hu_2001}
Hu W.,  2001, \mn@doi [Phys. Rev. D] {10.1103/PhysRevD.64.083005}, 64, 083005

\bibitem[\protect\citeauthoryear{Hu \& Okamoto}{Hu \& Okamoto}{2002}]{Hu_2002}
Hu W.,  Okamoto T.,  2002, \mn@doi [The Astrophysical Journal]
  {10.1086/341110}, 574, 566

\bibitem[\protect\citeauthoryear{Hunter}{Hunter}{2007}]{Hunter:2007}
Hunter J.~D.,  2007, \mn@doi [Computing In Science \& Engineering]
  {10.1109/MCSE.2007.55}, 9, 90

\bibitem[\protect\citeauthoryear{Jones, Oliphant, Peterson  et~al.}{Jones
  et~al.}{01  }]{scipy}
Jones E.,  Oliphant T.,  Peterson P.,   et~al., 2001--, {SciPy}: Open source
  scientific tools for {Python}, \url {http://www.scipy.org/}

\bibitem[\protect\citeauthoryear{Knox, Scoccimarro  \& Dodelson}{Knox
  et~al.}{1998}]{Knox:1998fp}
Knox L.,  Scoccimarro R.,   Dodelson S.,  1998, \mn@doi [Phys. Rev. Lett.]
  {10.1103/PhysRevLett.81.2004}, 81, 2004

\bibitem[\protect\citeauthoryear{Kulkarni, Choudhury, Puchwein  \&
  Haehnelt}{Kulkarni et~al.}{2017}]{Kulkarni:2017qwu}
Kulkarni G.,  Choudhury T.~R.,  Puchwein E.,   Haehnelt M.~G.,  2017, \mn@doi
  [Mon. Not. Roy. Astron. Soc.] {10.1093/mnras/stx1167}, 469, 4283

\bibitem[\protect\citeauthoryear{{La Plante}, Battaglia, Natarajan, Peterson,
  Trac, Cen  \& Loeb}{{La Plante} et~al.}{2014}]{LaPlante:2014}
{La Plante} P.,  Battaglia N.,  Natarajan A.,  Peterson J.~B.,  Trac H.,  Cen
  R.,   Loeb A.,  2014, \apj, \href
  {http://stacks.iop.org/0004-637X/789/i=1/a=31} {789, 31}

\bibitem[\protect\citeauthoryear{Lewis, Challinor  \& Lasenby}{Lewis
  et~al.}{2000}]{Lewis:1999bs}
Lewis A.,  Challinor A.,   Lasenby A.,  2000, \mn@doi [Astrophys. J.]
  {10.1086/309179}, 538, 473

\bibitem[\protect\citeauthoryear{{Limber}}{{Limber}}{1953}]{Limber_1953}
{Limber} D.~N.,  1953, \mn@doi [\apj] {10.1086/145672}, \href
  {https://ui.adsabs.harvard.edu/abs/1953ApJ...117..134L} {117, 134}

\bibitem[\protect\citeauthoryear{{Ma} \& {Fry}}{{Ma} \&
  {Fry}}{2002}]{Ma_Fry_2002}
{Ma} C.-P.,  {Fry} J.~N.,  2002, \mn@doi [\prl]
  {10.1103/PhysRevLett.88.211301}, \href
  {https://ui.adsabs.harvard.edu/abs/2002PhRvL..88u1301M} {88, 211301}

\bibitem[\protect\citeauthoryear{{Ma}, {Helgason}, {Komatsu}, {Ciardi}  \&
  {Ferrara}}{{Ma} et~al.}{2018}]{Ma_2018}
{Ma} Q.,  {Helgason} K.,  {Komatsu} E.,  {Ciardi} B.,   {Ferrara} A.,  2018,
  \mn@doi [\mnras] {10.1093/mnras/sty543}, \href
  {https://ui.adsabs.harvard.edu/abs/2018MNRAS.476.4025M} {476, 4025}

\bibitem[\protect\citeauthoryear{{McQuinn}, {Furlanetto}, {Hernquist}, {Zahn}
  \& {Zaldarriaga}}{{McQuinn} et~al.}{2005}]{Mcquinn_2005}
{McQuinn} M.,  {Furlanetto} S.~R.,  {Hernquist} L.,  {Zahn} O.,   {Zaldarriaga}
  M.,  2005, \mn@doi [\apj] {10.1086/432049}, \href
  {https://ui.adsabs.harvard.edu/abs/2005ApJ...630..643M} {630, 643}

\bibitem[\protect\citeauthoryear{{Mellema} et~al.,}{{Mellema}
  et~al.}{2013}]{2013ExA....36..235M}
{Mellema} G.,  et~al., 2013, \mn@doi [Experimental Astronomy]
  {10.1007/s10686-013-9334-5}, \href
  {https://ui.adsabs.harvard.edu/abs/2013ExA....36..235M} {36, 235}

\bibitem[\protect\citeauthoryear{{Mellema}, {Koopmans}, {Shukla}, {Datta},
  {Mesinger}  \& {Majumdar}}{{Mellema} et~al.}{2015}]{2015aska.confE..10M}
{Mellema} G.,  {Koopmans} L.,  {Shukla} H.,  {Datta} K.~K.,  {Mesinger} A.,
  {Majumdar} S.,  2015, in Advancing Astrophysics with the Square Kilometre
  Array (AASKA14). p.~10 (\mn@eprint {arXiv} {1501.04203})

\bibitem[\protect\citeauthoryear{Mesinger \& Furlanetto}{Mesinger \&
  Furlanetto}{2007}]{Mesinger:2007}
Mesinger A.,  Furlanetto S.,  2007, \apj, \href
  {http://stacks.iop.org/0004-637X/669/i=2/a=663} {669, 663}

\bibitem[\protect\citeauthoryear{{Mesinger}, {Furlanetto}  \& {Cen}}{{Mesinger}
  et~al.}{2011}]{2011MNRAS.411..955M}
{Mesinger} A.,  {Furlanetto} S.,   {Cen} R.,  2011, \mn@doi [\mnras]
  {10.1111/j.1365-2966.2010.17731.x}, \href
  {http://adsabs.harvard.edu/abs/2011MNRAS.411..955M} {411, 955}

\bibitem[\protect\citeauthoryear{Mesinger, McQuinn  \& Spergel}{Mesinger
  et~al.}{2012}]{Mesinger_2012}
Mesinger A.,  McQuinn M.,   Spergel D.~N.,  2012, \mn@doi [Monthly Notices of
  the Royal Astronomical Society] {10.1111/j.1365-2966.2012.20713.x}, 422, 1403

\bibitem[\protect\citeauthoryear{{Mondal}, {Bharadwaj}  \& {Datta}}{{Mondal}
  et~al.}{2018}]{2018MNRAS.474.1390M}
{Mondal} R.,  {Bharadwaj} S.,   {Datta} K.~K.,  2018, \mn@doi [\mnras]
  {10.1093/mnras/stx2888}, \href
  {https://ui.adsabs.harvard.edu/abs/2018MNRAS.474.1390M} {474, 1390}

\bibitem[\protect\citeauthoryear{Mortonson \& Hu}{Mortonson \&
  Hu}{2007}]{Mortonson:2007}
Mortonson M.~J.,  Hu W.,  2007, \apj, 657, 1

\bibitem[\protect\citeauthoryear{Mukherjee, Paul  \& Choudhury}{Mukherjee
  et~al.}{2019}]{Mukherjee_2019}
Mukherjee S.,  Paul S.,   Choudhury T.~R.,  2019, \mn@doi [Monthly Notices of
  the Royal Astronomical Society] {10.1093/mnras/stz1002}, 486, 2042

\bibitem[\protect\citeauthoryear{{Nozawa}, {Itoh}  \& {Kohyama}}{{Nozawa}
  et~al.}{1998}]{Nozawa:1998}
{Nozawa} S.,  {Itoh} N.,   {Kohyama} Y.,  1998, \mn@doi [\apj]
  {10.1086/306401}, \href {http://adsabs.harvard.edu/abs/1998ApJ...508...17N}
  {508, 17}

\bibitem[\protect\citeauthoryear{{Ostriker} \& {Vishniac}}{{Ostriker} \&
  {Vishniac}}{1986}]{1986ApJ...306L..51O}
{Ostriker} J.~P.,  {Vishniac} E.~T.,  1986, \mn@doi [\apjl] {10.1086/184704},
  \href {https://ui.adsabs.harvard.edu/abs/1986ApJ...306L..51O} {306, L51}

\bibitem[\protect\citeauthoryear{Paciga et~al.,}{Paciga
  et~al.}{2013}]{Paciga_2013}
Paciga G.,  et~al., 2013, \mn@doi [Monthly Notices of the Royal Astronomical
  Society] {10.1093/mnras/stt753}, 433, 639

\bibitem[\protect\citeauthoryear{Park, Shapiro, Komatsu, Iliev, Ahn  \&
  Mellema}{Park et~al.}{2013}]{Park_2013}
Park H.,  Shapiro P.~R.,  Komatsu E.,  Iliev I.~T.,  Ahn K.,   Mellema G.,
  2013, \mn@doi [The Astrophysical Journal] {10.1088/0004-637x/769/2/93}, 769,
  93

\bibitem[\protect\citeauthoryear{{Park}, {Komatsu}, {Shapiro}, {Koda}  \&
  {Mao}}{{Park} et~al.}{2016}]{Park_2016}
{Park} H.,  {Komatsu} E.,  {Shapiro} P.~R.,  {Koda} J.,   {Mao} Y.,  2016,
  \mn@doi [\apj] {10.3847/0004-637X/818/1/37}, \href
  {https://ui.adsabs.harvard.edu/abs/2016ApJ...818...37P} {818, 37}

\bibitem[\protect\citeauthoryear{Parsons et~al.,}{Parsons
  et~al.}{2014}]{Parsons_2014}
Parsons A.~R.,  et~al., 2014, \mn@doi [The Astrophysical Journal]
  {10.1088/0004-637x/788/2/106}, 788, 106

\bibitem[\protect\citeauthoryear{P\'erez \& Granger}{P\'erez \&
  Granger}{2007}]{PER-GRA:2007}
P\'erez F.,  Granger B.~E.,  2007, \mn@doi [Computing in Science and
  Engineering] {10.1109/MCSE.2007.53}, 9, 21

\bibitem[\protect\citeauthoryear{{Planck Collaboration} et~al.,}{{Planck
  Collaboration} et~al.}{2014}]{Planck:2014}
{Planck Collaboration} et~al., 2014, \mn@doi [\aap]
  {10.1051/0004-6361/201321591}, \href
  {https://ui.adsabs.harvard.edu/\#abs/2014A&A...571A..16P} {571, A16}

\bibitem[\protect\citeauthoryear{{Planck Collaboration} et~al.,}{{Planck
  Collaboration} et~al.}{2018}]{Planck:2018}
{Planck Collaboration} et~al., 2018, arXiv:1807.06209, \href
  {http://adsabs.harvard.edu/abs/2018arXiv180706209P} {}

\bibitem[\protect\citeauthoryear{{Reichardt} et~al.,}{{Reichardt}
  et~al.}{2012}]{2012ApJ...755...70R}
{Reichardt} C.~L.,  et~al., 2012, \mn@doi [\apj] {10.1088/0004-637X/755/1/70},
  \href {https://ui.adsabs.harvard.edu/abs/2012ApJ...755...70R} {755, 70}

\bibitem[\protect\citeauthoryear{{Reichardt} et~al.,}{{Reichardt}
  et~al.}{2020}]{2020arXiv200206197R}
{Reichardt} C.~L.,  et~al., 2020, arXiv e-prints, \href
  {https://ui.adsabs.harvard.edu/abs/2020arXiv200206197R} {p. arXiv:2002.06197}

\bibitem[\protect\citeauthoryear{Roy, Kulkarni, Meerburg, Challinor,
  Baccigalupi, Lapi  \& Haehnelt}{Roy et~al.}{2020}]{Roy:2020cqn}
Roy A.,  Kulkarni G.,  Meerburg P.~D.,  Challinor A.,  Baccigalupi C.,  Lapi
  A.,   Haehnelt M.~G.,  2020, arXiv:2004.02927

\bibitem[\protect\citeauthoryear{{Salvaterra}, {Ciardi}, {Ferrara}  \&
  {Baccigalupi}}{{Salvaterra} et~al.}{2005}]{2005MNRAS.360.1063S}
{Salvaterra} R.,  {Ciardi} B.,  {Ferrara} A.,   {Baccigalupi} C.,  2005,
  \mn@doi [\mnras] {10.1111/j.1365-2966.2005.09089.x}, \href
  {https://ui.adsabs.harvard.edu/abs/2005MNRAS.360.1063S} {360, 1063}

\bibitem[\protect\citeauthoryear{Santos, Cooray, Haiman, Knox  \& Ma}{Santos
  et~al.}{2003}]{Santos_2003}
Santos M.~G.,  Cooray A.,  Haiman Z.,  Knox L.,   Ma C.-P.,  2003, \mn@doi [The
  Astrophysical Journal] {10.1086/378772}, 598, 756

\bibitem[\protect\citeauthoryear{Santos, Amblard, Pritchard, Trac, Cen  \&
  Cooray}{Santos et~al.}{2008}]{Santos:2007dn}
Santos M.~G.,  Amblard A.,  Pritchard J.,  Trac H.,  Cen R.,   Cooray A.,
  2008, \mn@doi [Astrophys. J.] {10.1086/592487}, 689, 1

\bibitem[\protect\citeauthoryear{Seehars, Paranjape, Witzemann, Refregier,
  Amara  \& Akeret}{Seehars et~al.}{2016}]{Seehars:2015ada}
Seehars S.,  Paranjape A.,  Witzemann A.,  Refregier A.,  Amara A.,   Akeret
  J.,  2016, \mn@doi [JCAP] {10.1088/1475-7516/2016/03/001}, 1603, 001

\bibitem[\protect\citeauthoryear{Sehgal et~al.}{Sehgal
  et~al.}{2020}]{Sehgal:2020yja}
Sehgal N.,  et~al., 2020

\bibitem[\protect\citeauthoryear{Seljak}{Seljak}{1998}]{Seljak:1997ep}
Seljak U.,  1998, \mn@doi [Astrophys. J.] {10.1086/306225}, 506, 64

\bibitem[\protect\citeauthoryear{{Shaw}, {Rudd}  \& {Nagai}}{{Shaw}
  et~al.}{2012}]{2012ApJ...756...15S}
{Shaw} L.~D.,  {Rudd} D.~H.,   {Nagai} D.,  2012, \mn@doi [\apj]
  {10.1088/0004-637X/756/1/15}, \href
  {https://ui.adsabs.harvard.edu/abs/2012ApJ...756...15S} {756, 15}

\bibitem[\protect\citeauthoryear{{Sheth} \& {Tormen}}{{Sheth} \&
  {Tormen}}{2002}]{2002MNRAS.329...61S}
{Sheth} R.~K.,  {Tormen} G.,  2002, \mn@doi [\mnras]
  {10.1046/j.1365-8711.2002.04950.x}, \href
  {https://ui.adsabs.harvard.edu/abs/2002MNRAS.329...61S} {329, 61}

\bibitem[\protect\citeauthoryear{Sievers et~al.}{Sievers
  et~al.}{2013}]{Sievers:2013ica}
Sievers J.~L.,  et~al., 2013, \mn@doi [JCAP] {10.1088/1475-7516/2013/10/060},
  10, 060

\bibitem[\protect\citeauthoryear{Smith \& Ferraro}{Smith \&
  Ferraro}{2017}]{Smith:2016lnt}
Smith K.~M.,  Ferraro S.,  2017, \mn@doi [Phys. Rev. Lett.]
  {10.1103/PhysRevLett.119.021301}, 119, 021301

\bibitem[\protect\citeauthoryear{{Springel}}{{Springel}}{2005}]{GADGET2:2005}
{Springel} V.,  2005, \mn@doi [\mnras] {10.1111/j.1365-2966.2005.09655.x},
  \href {http://adsabs.harvard.edu/abs/2005MNRAS.364.1105S} {364, 1105}

\bibitem[\protect\citeauthoryear{{Su}, {Yadav}, {McQuinn}, {Yoo}  \&
  {Zaldarriaga}}{{Su} et~al.}{2011}]{2011arXiv1106.4313S}
{Su} M.,  {Yadav} A. P.~S.,  {McQuinn} M.,  {Yoo} J.,   {Zaldarriaga} M.,
  2011, arXiv e-prints, \href
  {https://ui.adsabs.harvard.edu/\#abs/2011arXiv1106.4313S} {p.
  arXiv:1106.4313}

\bibitem[\protect\citeauthoryear{{Sunyaev} \& {Zeldovich}}{{Sunyaev} \&
  {Zeldovich}}{1970}]{Sunyaev_1970}
{Sunyaev} R.~A.,  {Zeldovich} Y.~B.,  1970, \mn@doi [\apss]
  {10.1007/BF00653471}, \href
  {https://ui.adsabs.harvard.edu/abs/1970Ap%26SS...7....3S} {7, 3}

\bibitem[\protect\citeauthoryear{{Sunyaev} \& {Zeldovich}}{{Sunyaev} \&
  {Zeldovich}}{1980}]{Sunyaev_1980}
{Sunyaev} R.~A.,  {Zeldovich} I.~B.,  1980, \mn@doi [\araa]
  {10.1146/annurev.aa.18.090180.002541}, \href
  {https://ui.adsabs.harvard.edu/abs/1980ARA%26A..18..537S} {18, 537}

\bibitem[\protect\citeauthoryear{{Suzuki} et~al.,}{{Suzuki}
  et~al.}{2018}]{2018JLTP..193.1048S}
{Suzuki} A.,  et~al., 2018, \mn@doi [Journal of Low Temperature Physics]
  {10.1007/s10909-018-1947-7}, \href
  {https://ui.adsabs.harvard.edu/abs/2018JLTP..193.1048S} {193, 1048}

\bibitem[\protect\citeauthoryear{{Thornton} et~al.,}{{Thornton}
  et~al.}{2016}]{2016ApJS..227...21T}
{Thornton} R.~J.,  et~al., 2016, \mn@doi [\apjs] {10.3847/1538-4365/227/2/21},
  \href {https://ui.adsabs.harvard.edu/abs/2016ApJS..227...21T} {227, 21}

\bibitem[\protect\citeauthoryear{Tingay et~al.,}{Tingay
  et~al.}{2013}]{tingay_2013}
Tingay S.~J.,  et~al., 2013, \mn@doi [Publications of the Astronomical Society
  of Australia] {10.1017/pasa.2012.007}, 30, e007

\bibitem[\protect\citeauthoryear{{Van Haarlem, M. P.} et~al.,}{{Van Haarlem, M.
  P.} et~al.}{2013}]{Haarlem_2013}
{Van Haarlem, M. P.} et~al., 2013, \mn@doi [A\&A]
  {10.1051/0004-6361/201220873}, 556, A2

\bibitem[\protect\citeauthoryear{{Wyithe} \& {Loeb}}{{Wyithe} \&
  {Loeb}}{2003}]{Wyithe_2003}
{Wyithe} J.~S.~B.,  {Loeb} A.,  2003, \mn@doi [\apjl] {10.1086/375682}, \href
  {https://ui.adsabs.harvard.edu/abs/2003ApJ...588L..69W} {588, L69}

\bibitem[\protect\citeauthoryear{{Zahn}, {Zaldarriaga}, {Hernquist}  \&
  {McQuinn}}{{Zahn} et~al.}{2005}]{Zahn_2005}
{Zahn} O.,  {Zaldarriaga} M.,  {Hernquist} L.,   {McQuinn} M.,  2005, \mn@doi
  [\apj] {10.1086/431947}, \href
  {https://ui.adsabs.harvard.edu/abs/2005ApJ...630..657Z} {630, 657}

\bibitem[\protect\citeauthoryear{Zahn, Lidz, McQuinn, Dutta, Hernquist,
  Zaldarriaga  \& Furlanetto}{Zahn et~al.}{2007}]{Zahn:2007}
Zahn O.,  Lidz A.,  McQuinn M.,  Dutta S.,  Hernquist L.,  Zaldarriaga M.,
  Furlanetto S.~R.,  2007, \apj, \href
  {http://stacks.iop.org/0004-637X/654/i=1/a=12} {654, 12}

\bibitem[\protect\citeauthoryear{Zahn et~al.,}{Zahn et~al.}{2012}]{Zahn_2012}
Zahn O.,  et~al., 2012, \mn@doi [The Astrophysical Journal]
  {10.1088/0004-637x/756/1/65}, 756, 65

\bibitem[\protect\citeauthoryear{Zaldarriaga \& Seljak}{Zaldarriaga \&
  Seljak}{1998}]{Zaldarriaga:1998ar}
Zaldarriaga M.,  Seljak U.,  1998, \mn@doi [Phys. Rev. D]
  {10.1103/PhysRevD.58.023003}, 58, 023003

\bibitem[\protect\citeauthoryear{Zhang, Zheng  \& Jing}{Zhang
  et~al.}{2015}]{Zhang:2014hra}
Zhang P.,  Zheng Y.,   Jing Y.,  2015, \mn@doi [Phys. Rev. D]
  {10.1103/PhysRevD.91.043522}, 91, 043522

\bibitem[\protect\citeauthoryear{Zheng, Zhang  \& Jing}{Zheng
  et~al.}{2015}]{Zheng:2014ywa}
Zheng Y.,  Zhang P.,   Jing Y.,  2015, \mn@doi [Phys. Rev. D]
  {10.1103/PhysRevD.91.043523}, 91, 043523

\bibitem[\protect\citeauthoryear{{van der Walt}, {Colbert}  \&
  {Varoquaux}}{{van der Walt} et~al.}{2011}]{2011CSE....13b..22V}
{van der Walt} S.,  {Colbert} S.~C.,   {Varoquaux} G.,  2011, \mn@doi
  [Computing in Science and Engineering] {10.1109/MCSE.2011.37}, \href
  {https://ui.adsabs.harvard.edu/abs/2011CSE....13b..22V} {13, 22}

\makeatother
\end{thebibliography}

\appendix
\section{New scaling relation for the Kinematic SZ and CMB $B$-mode polarization signal}\label{scal}

\begin{figure}
\centering
{\includegraphics[trim={0cm 0cm 0cm 0cm},clip,width=0.5\textwidth]{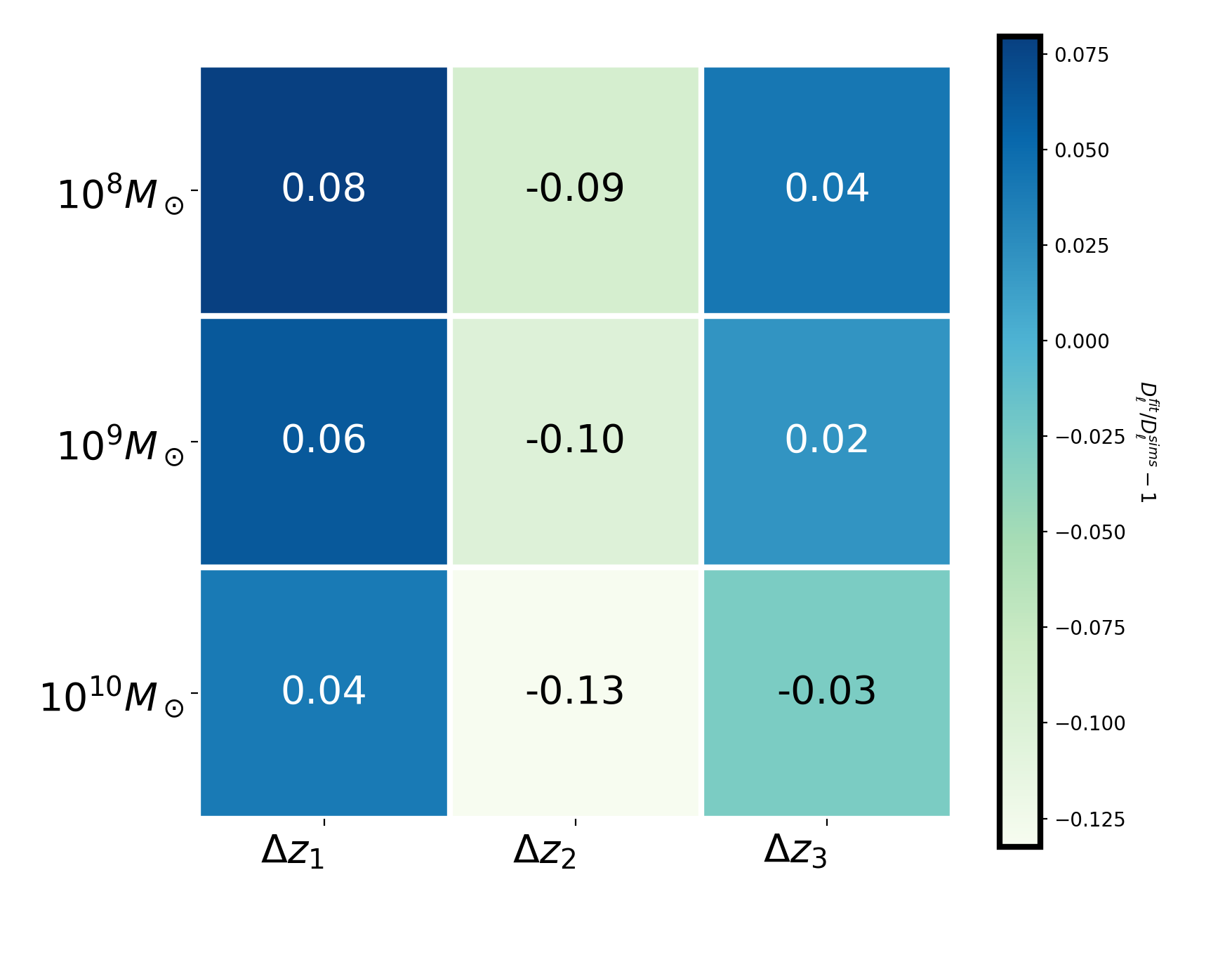}}
\caption{We show the relative difference ($D^{fit}_l/D^{sims}_l -1$) of the kSZ power spectrum between the scaling relation and simulation results for the cases with optical depth $\tau=0.054$. The maximum departure is around $13\%$.  The deviations for $\tau=0.061$ are also similar to this case.}
\label{tau54_res}
\end{figure}

\begin{figure}
\centering
{\includegraphics[trim={0cm 0cm 0cm 0cm},clip,width=0.5\textwidth]{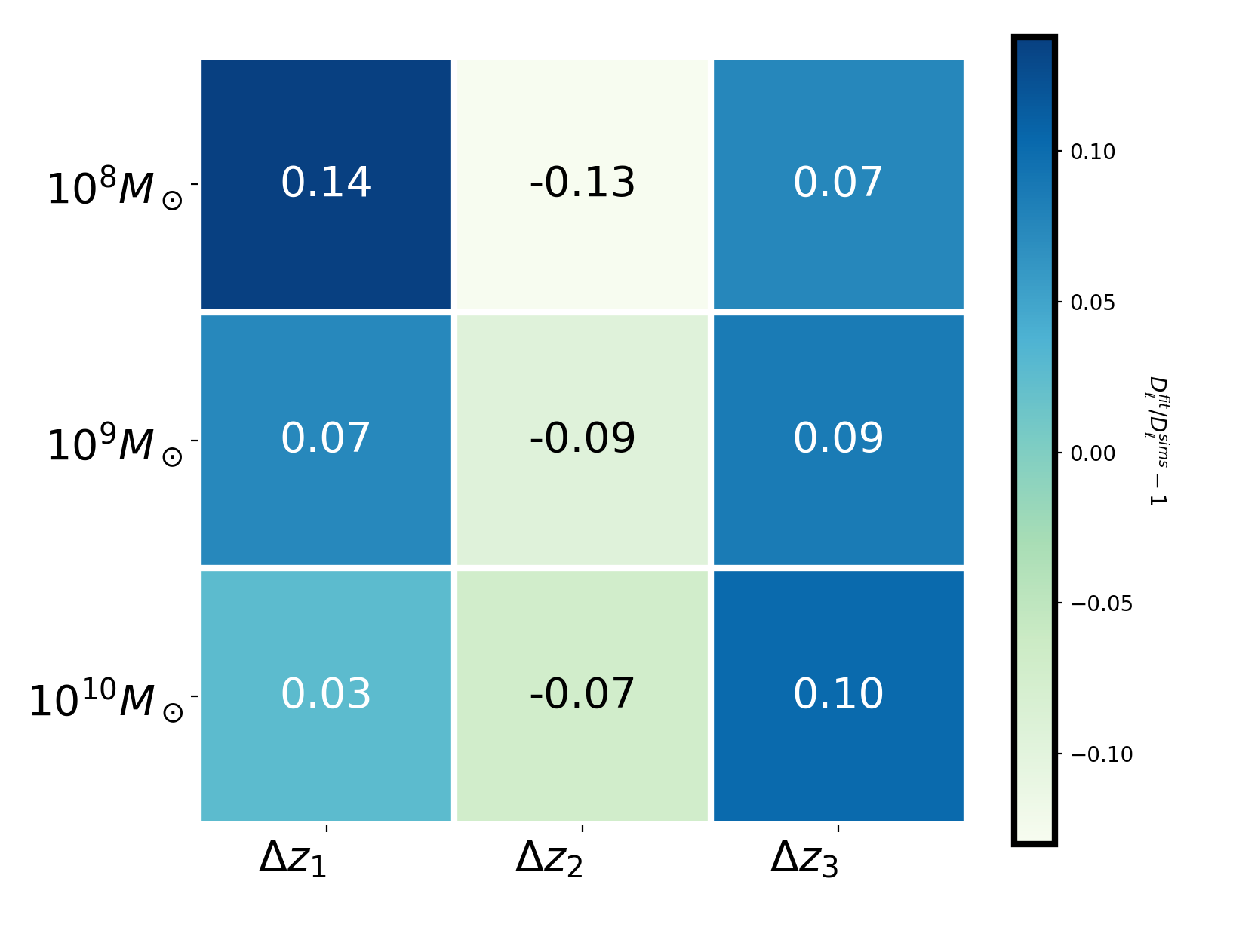}}
\caption{We show the relative difference ($D^{fit}_l/D^{sims}_l -1$) of the B-mode power spectrum between the scaling relation and simulation results for the cases with optical depth $\tau=0.054$. The maximum departure is around $14\%$. The deviations for $\tau=0.061$ are also similar to this case.}
\label{tau54_resBB}
\end{figure}

\begin{figure*}
\centering
{\includegraphics[trim={0cm 0cm 0cm 0cm},clip,width=1.0\textwidth]{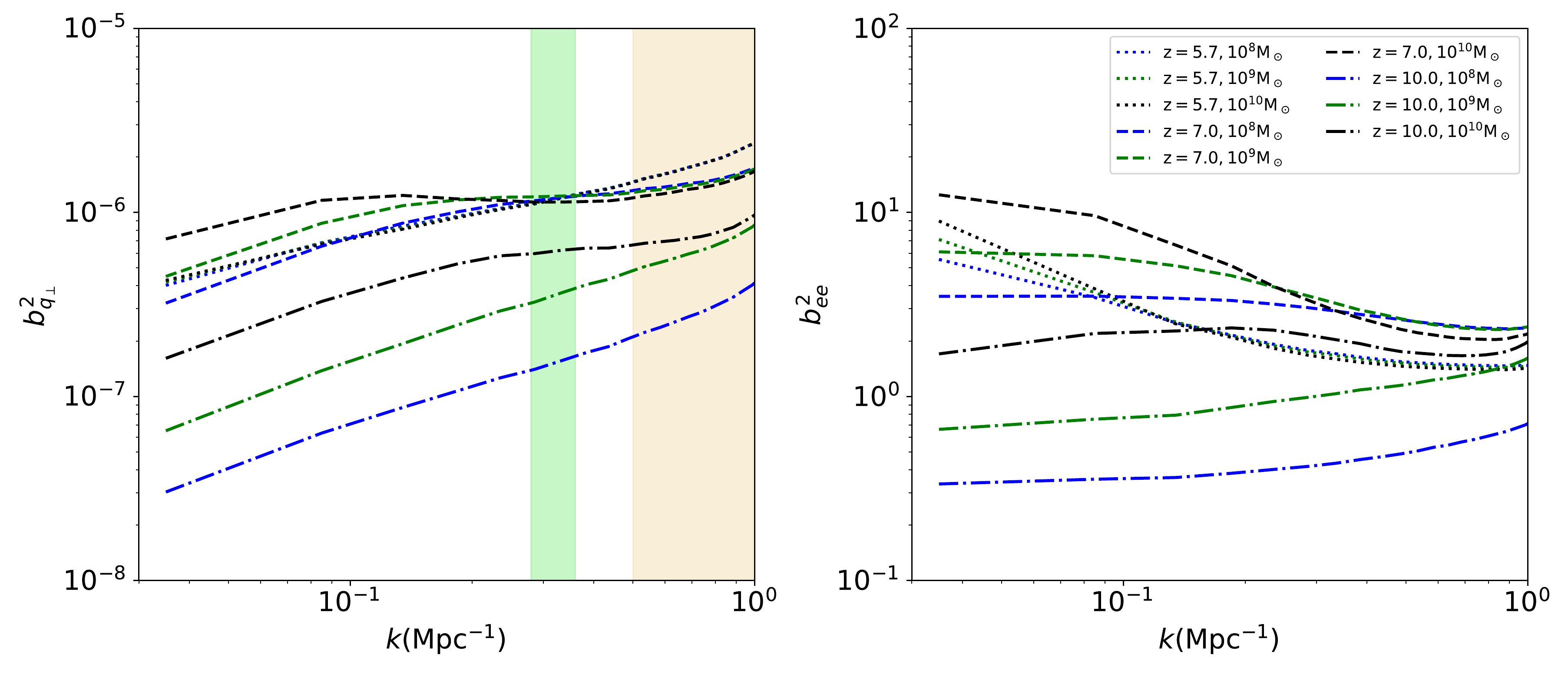}}
\caption{Dependence of the bias parameters $b_{q_{\perp}}$ and $b_{ee}$ with $k$ at redshifts $z=5.7, 7$ and $10$ which correspond to $\rm{Q_{HII}} \approx 0.9, 0.5$ and $0.1$ respectively for the fiducial reionization history. In $b_{q_{\perp}}$, the region shaded in green denotes the modes which contributes to the kSZ power spectrum $C_{l}^{\rm{kSZ}}$ at $l=3000$ for the reionization histories considered in this analysis. 
The region shaded in 
in yellow ($k\geq0.5$ Mpc$^{-1}$) is possibly affected by finite sampling artefacts \citep{Zhang:2014hra, Zheng:2014ywa}.
}
\label{bias_params}
\end{figure*}

Simulation-based study of cosmic reionization shows that the spatial fluctuations in electron density can play a crucial role in understanding the amplitude of the kSZ and $B$-mode polarization signal. So to connect the amplitude of the kSZ signal and the $B$-mode polarization with the parameters related to the EoR, new scaling relations will be useful. 

We model the amplitude of the angular power spectrum ($D^X_l\equiv l(l+1)C^X_l/2\pi$) by a parametric form 
\begin{equation}
D^{X}_{l=l_X}= A_X(\tau) (\Delta z)^{\alpha_X}\left(b^2_X(l=l_X)\right)^{\beta_X},
\end{equation} 
where, $X \in \{kSZ, BB\}$, $\tau$ is the CMB optical optical depth, $\Delta z$ is the duration of the EoR, and $b_X$ are the bias parameters defined with respect to the DM matter power spectrum defined in equations (\ref{bkSZ}) and (\ref{bBB}). The choice of writing the kSZ and $B$-mode polarization amplitudes in terms of the bias parameter $b_X(l)$ makes it possible to avoid the dependence on the overall normalization of the matter power spectrum in the scaling relation. We use $l_{\rm kSZ} = 3000$ and $l_{BB} = 200$.

Using the set of $18$ simulations (which are discussed in Sec. \ref{Simulation}) for different cases (by varying $\tau$, $\Delta z$, and minimum halo mass $M_{\rm min}$), we calculate the value of $b_{X}(l)$ and the value of $D_{l=l_X}$, and obtain the best-fit parameters for $A(\tau)$, $\alpha$, and $\beta$ by minimizing the $\chi^2$ defined as 
\begin{equation}
\chi^2= \sum_i \left(\frac{D^{X}_{l, i} - A_X(\tau) (\Delta z_i)^{\alpha_X} \left(b^2_{X, i}(l)\right)^{\beta_X}} {\Sigma_{l,i}^{X}}\right)^2,
\end{equation}
where the index $i$ denotes the individual simulation cases, and $\left(\Sigma_{l,i}^{X}\right)^2$ is the variance in the kSZ/$B$-mode polarization  amplitude which is taken as the cosmic variance value defined as $\Sigma^{X}_{l,i}= \sqrt{\frac{2}{2l+1}}D^{X}_{l,i}$. 

The corresponding best-fit parameter leads to a new scaling relation for the kSZ amplitude given in equation (\ref{newscaling}). The maximum difference between the scaling relation and the simulation results are $\sim 13\%$ as shown in Figure~\ref{tau54_res} for the cases with $\tau= 0.054$. We found that the departure is also similar for cases with $\tau=0.061$.

Similarly, for the $B$-mode polarization amplitude, we obtained the scaling relation as given in equation (\ref{newscalingbmode}). The maximum difference between the scaling relation and simulation results in this case are $\sim 14\%$ as shown in Figure~\ref{tau54_resBB}, and is similar for both $\tau= 0.054$ and $\tau= 0.061$. 

In Figure~\ref{bias_params}, we show the variation of the bias parameters $b_{q_{\perp}}$ and $b_{ee}$ with $k$ at three stages of reionization for our fiducial case. 
In the green shaded region, we show the modes which contribute to the kSZ power spectrum at the CMB multipole $l=3000$ for all the reionization histories considered in the analysis.
Interestingly, we find that the simplified relation often used in the literature \citep[see, e.g.,][]{Ma_Fry_2002,Mcquinn_2005} $P_{q_{\perp}} = \langle v^2 \rangle~P_{ee} / 3$, where $\langle v^2 \rangle$ is the rms of the peculiar velocity field, does \emph{not} hold at scales relevant for the kSZ signal. This is mainly because the connected fourth moment of the ionization and velocity fluctuations, which is ignored in the simplified relation, becomes non-negligible at these scales \citep{Park_2016}. The amplitude of this term is driven by the non-linearities in the velocity field and patchiness in the ionization field and hence becomes particularly important at lower redshifts and for higher values of $M_{\mathrm{min}}$.

\label{lastpage}

\end{document}